\documentclass{IEEEtran}
\ifCLASSINFOpdf
\else
\fi
\usepackage[justification=centering]{caption}
\usepackage{cuted}
\usepackage{amsmath}
\allowdisplaybreaks[1]
\usepackage{times,amsmath,epsfig}
\usepackage[latin1]{inputenc}
\usepackage{fancyhdr}
\usepackage{amsmath}
\usepackage{amsfonts}
\usepackage{amssymb}
\usepackage{mathrsfs}
\usepackage{float}
\usepackage{array}
\usepackage{graphicx}
\usepackage{epstopdf}
\usepackage{url}
\usepackage{subfigure}
\usepackage{bm}
\usepackage{breqn}
\usepackage{xcolor}
\usepackage{soul}
\usepackage{picinpar,graphicx}
\usepackage{cite}
\usepackage{multirow}
\usepackage{algorithm}
\usepackage{algpseudocode}
\usepackage{amsmath,lipsum}
\usepackage{multicol}

\usepackage[normalem]{ulem}

\begin{document}
\title{Symbol-level Integrated Sensing and Communication enabled Multiple Base Stations Cooperative Sensing}
\author{Zhiqing Wei,~\IEEEmembership{Member, IEEE,}
	Ruizhong Xu,~
	Zhiyong Feng,~\IEEEmembership{Senior Member, IEEE,}\\
	Huici Wu,~\IEEEmembership{Member, IEEE,}
	Ning Zhang,~\IEEEmembership{Senior Member, IEEE,}\\
	Wangjun Jiang,~\IEEEmembership{Student Member, IEEE,}
	Xiaoyu Yang,~\IEEEmembership{Student Member, IEEE}
	
	\thanks{Zhiqing Wei, Ruizhong Xu, Zhiyong Feng, Wangjun Jiang and Xiaoyu Yang 
		are with the Key Laboratory of Universal Wireless Communications, Ministry of Education, 
		School of Information and Communication Engineering, 
		Beijing University of Posts and Telecommunications, Beijing 100876, China 
		(emails: weizhiqing@bupt.edu.cn; xuruizhong@bupt.edu.cn; fengzy@bupt.edu.cn; jiangwangjun@bupt.edu.cn;  xiaoyu.yang@bupt.edu.cn).
		
		Huici Wu is with the National Engineering Lab for Mobile Network Technologies, Beijing University of Posts and Telecommunications, Beijing 100876, China (email: dailywu@bupt.edu.cn). 
		
		Ning Zhang is with the Department of Electrical and Computer Engineering, 
		University of Windsor, Windsor, ON, N9B 3P4,
		Canada (email: ning.zhang@uwindsor.ca).}}

\maketitle

\begin{abstract}
With the support of integrated sensing and communication (ISAC) technology,
mobile communication system will integrate the function of wireless sensing,
thereby facilitating new intelligent applications such as smart city and intelligent transportation.
Due to the limited sensing accuracy and sensing range of single base station (BS),
multi-BS cooperative sensing can be applied to realize high-accurate,
long-range and continuous sensing,
exploiting the specific advantages of large-scale networked mobile communication system.
This paper proposes a cooperative sensing method suitable to mobile communication systems,
which applies symbol-level sensing information fusion 
to estimate the location and velocity of target.
With the demodulation symbols obtained from the echo signals of multiple BSs, 
the phase features
contained in the demodulation symbols are used in the fusion procedure, 
which realizes cooperative sensing with the synchronization level of mobile communication system. 
Compared with the signal-level fusion in the area of distributed aperture
coherence-synthetic radars, the requirement of synchronization is much lower.
When signal-to-noise ratio (SNR) is --5 dB,
it is evaluated that symbol-level multi-BS cooperative sensing 
effectively improves the accuracy of distance and velocity estimation of target.
Compared with single-BS sensing,
the accuracy of distance and velocity estimation is improved by 40\% and 72\%, respectively. 
Compared with data-level multi-BS cooperative sensing based on maximum likelihood (ML) estimation,
the accuracy of location and velocity estimation is improved by 12\% and 63\%, respectively. 
This work may provide a guideline for the design of multi-BS
cooperative sensing system to exploit the widely deployed
networked mobile communication system.
\end{abstract}


\IEEEpeerreviewmaketitle

\section{Introduction}
\begin{figure*}[!ht]
	\centering
	\subfigure[Multi-BS cooperative sensing to extend the sensing range]{\includegraphics[width=0.46\textwidth]{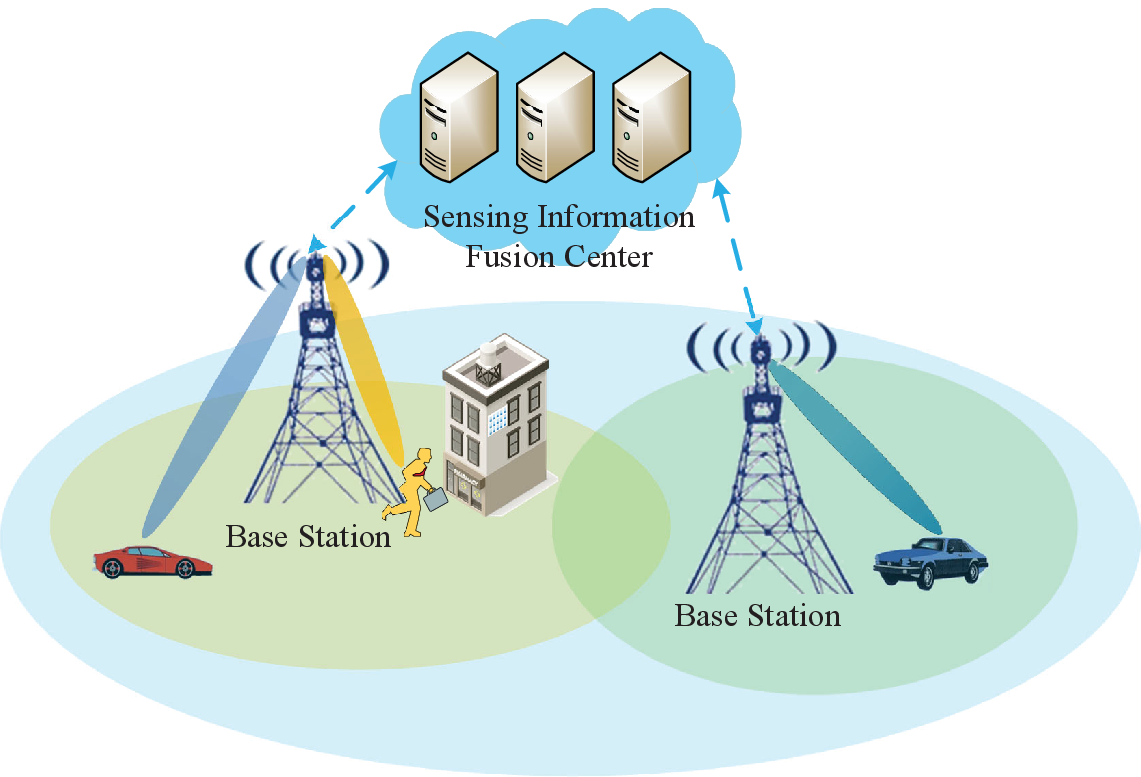}\label{1a}}
	\quad
	\subfigure[Multi-BS cooperative sensing to improve sensing accuracy]{\includegraphics[width=0.51\textwidth]{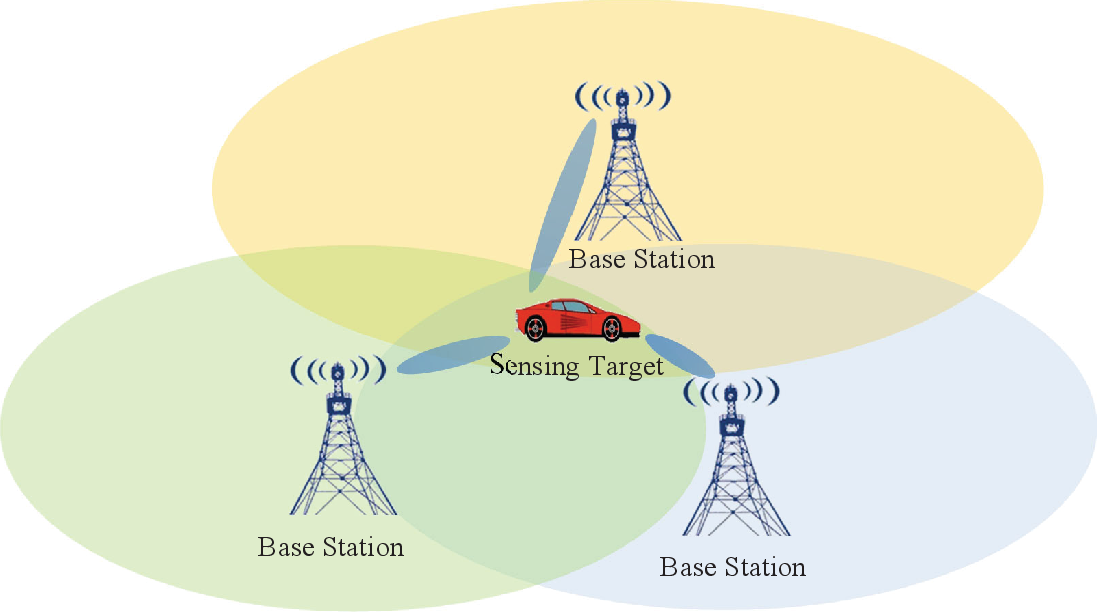}\label{1b}}
	\quad
	\subfigure[Multi-BS cooperative sensing to improve the success probability of target sensing]{\includegraphics[width=0.63\textwidth]{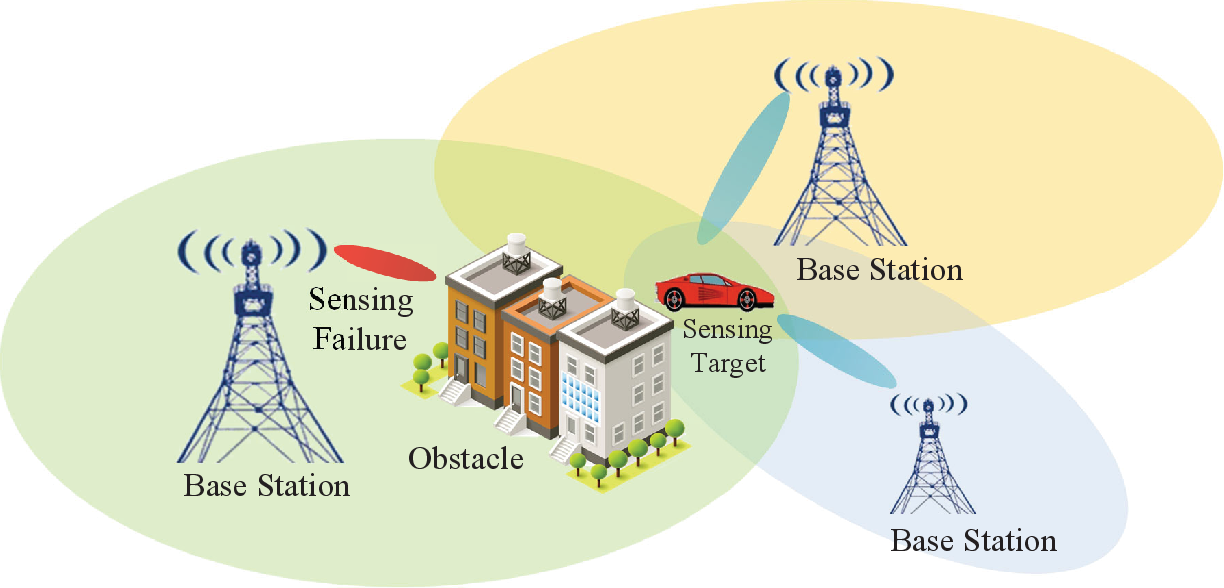}\label{1c}}
	\caption{Advantages of multi-BS cooperative sensing.}
	\label{fig1}
\end{figure*}
In the era of fifth-generationadvanced
(5G-A) and sixth-generation (6G) mobile communications,
new intelligent applications such as smart city,
intelligent transportation, and unmanned factory have emerged \cite{1}.
These intelligent applications urgently need the information infrastructure
with sensing and communication functions.
For example, environmental sensing is the essential function of autonomous
driving to realize safe and intelligent driving \cite{2,3}.
As a widely deployed information infrastructure, mobile communication system is
continuously integrating wireless sensing function with the assistance of
integrated sensing and communication (ISAC) technology \cite{1,4}, gradually evolving into
a unified information infrastructure that integrates sensing and communication capabilities
to support intelligent applications.

ISAC realizes radar sensing using the mobile communication system.
For the complex scenarios such as smart city and intelligent transportation,
the mobile communication system with ISAC capability needs to
have a large-coverage and high-accurate sensing capability.
However, single-BS sensing has insufficient accuracy for the distant targets \cite{5_0, 5}.
Therefore, multi-BS cooperative sensing
is promising to overcome
the limitation of single-BS sensing \cite{5_1}. 
Compared with single-BS sensing,
multi-BS cooperative sensing has the potential to improve sensing efficiency from the following aspects.
\begin{enumerate}
		\item \textbf{Extension of sensing range}: As shown in Fig. \ref{1a}, 
		multiple BSs transmit sensing results
		to the fusion center for sensing information fusion, extending the detection range \cite{5_0}.
		\item \textbf{Improvement of sensing accuracy}: As illustrated in Fig. \ref{1b},
		when a BS detects a distant target, 
		the sensing accuracy is limited due to the low signal-to-noise ratio
		(SNR) of the received echo signal. In this case, sensing accuracy 
		can be improved via multi-BS cooperative sensing \cite{6_0}.
		\item \textbf{Improvement of success probability of target sensing}:
		As shown in Fig. \ref{1c}, due to interference or obstacles,
		single-BS may fail to detect target.
		Through multi-BS cooperative sensing, multiple BSs
		can detect target from different positions and angles,
		which greatly improves the success probability of target sensing \cite{6_1}.
\end{enumerate}

Hence, multi-BS cooperative sensing expects to improve the sensing efficiency.
However, as the most similar research area of multi-BS cooperative sensing,
the existing cooperative sensing methods proposed for radar system
have extremely high requirements on
the synchronization accuracy between radars and the deployment locations of radars,
which cannot be directly applied in multi-BS cooperative sensing.
Therefore, the cooperative sensing method 
that is suitable to
mobile communication system needs to be designed.

Sensing information fusion is crucial for multi-BS/radar cooperative sensing,
which consists of data-level fusion and signal-level fusion.
In data-level fusion, the estimated results of target by multiple BSs/radars are 
fused for location and velocity estimation.
In signal-level fusion, the echo signals of multiple BSs/radars are fused 
for location and velocity estimation.
Data-level fusion methods can be further classified into
weighted average method \cite{6,7}, maximum likelihood (ML) method \cite{8,9,10,11,12,13},
least square method \cite{14,15}, and so on. 
In weighted average method,
the optimal weights for sensing information fusion are determined according to
the consistency and stability of the sensing data of multiple radars \cite{6,7}.
In ML method,
Prophet \emph{et al.} divided the detection area into multiple two-dimensional (2D) grids.
Then, the probability that target falls in each grid is derived to estimate 
the location of target \cite{8,9}.
Dash \emph{et al.} proposed a distributed ML estimation to detect and track target \cite{12}.
Weiss \emph{et al.} proposed an ML location estimation method for the stationary
transmitter whose delay and Doppler frequency shift information are
observed by moving receivers \cite{10}. Braun \emph{et al.} proposed
a velocity and distance estimation method for orthogonal frequency division multiplexing (OFDM)
radar by combining ML method
and fast Fourier transform (FFT) method \cite{11}.
Weiss \emph{et al.} proposed a direct position determination method,
locating target using the joint likelihood function of 
the echo signals of multiple targets \cite{13}.
As for least square method, Turlapaty \emph{et al.}
applied weighted OFDM waveform and least square method to detect mobile target \cite{14, 15}.
The data-level fusion does not have high requirement on synchronization
among multiple BSs and has low-complexity.
However, the sensing accuracy of data-level fusion is limited.

The signal-level fusion is applied in the distributed aperture coherence-synthetic radars,
where the airspace and energy resources of multi-aperture are applied
to improve the sensing range and accuracy \cite{16}.
The receive-coherence guarantees a high SNR gain. 
In contrast, full-coherence, which realizes transmit-coherence and receive-coherence simultaneously, 
will guarantee a higher SNR gain at the receiver \cite{17}.
However, full-coherence requires the setting of 
coherence parameters such as delay and phases according to the sensing results of target \cite{17}.
To solve this problem, 
Yang \emph{et al.} proposed the orthogonal signal design scheme to improve 
the estimation accuracy of
phase difference \cite{18}. 
Although the signal-level fusion for distributed aperture coherence-synthetic radars
has high sensing accuracy,
it requires extremely high synchronization accuracy (nanosecond level) and
optimized deployment positions of multiple radars,
which is not suitable to the mobile communication system.

Overall, the research on multi-BS cooperative sensing is still in its infancy.
Most of the cooperative sensing algorithms are proposed in the field of radar,
which have extremely high requirements on deployment locations of multiple radars
and time synchronization.
The cooperative sensing methods suitable to mobile communication systems are very rare.
In this paper, symbol-level cooperative sensing method is proposed for  
location and velocity estimation of target,
which is suitable to the synchronization level of mobile communication system. 
Traditional multi-node cooperative sensing methods operate on the target feature parameters such as distance, 
velocity uploaded by each node. However, the data used in the fusion operation of the proposed symbol-level 
multi-BS cooperative sensing method is the amplitude and phase information in the symbol components.
Compared with the signal-level fusion in the area of distributed aperture
coherence-synthetic radars, the requirement of synchronization is much lower.
Besides, the accuracy of location and velocity estimation is much 
higher than the data-level fusion.
In this paper, the multi-BS cooperative sensing method using communication signal, instead of the typical radar signals such as linear frequency modulated (LFM) signal and pulse signals, are proposed.
The detailed contributions of this work are listed as follows.

\begin{enumerate}
	\item The single-BS signal preprocessing method is proposed,
	where the symbol vectors that preserve the distance and radial velocity information
	are obtained with the phase adjustment and accumulation of symbols,
	reducing the data transmitted to the fusion center
	and the bandwidth requirement for multi-BS sensing information fusion.
	
	\item The phase calibration method for the symbol vectors of each BS is proposed
	to facilitate multi-BS sensing information fusion,
	where the phases change due to the distance or velocity of
	target is retained.
	
	\item The symbol-level multi-BS sensing information fusion is proposed,
	where the symbol vectors are compensated and cumulative multiplied,
	improving the SNR at the fusion center and the accuracy of target sensing.
\end{enumerate}

The structure of this paper is as follows.
The ISAC signal model is provided in Section II. 
A multi-BS sensing information fusion algorithm is proposed in Section III. Algorithm performance analysis is proposed in Section IV.
Simulation results are shown in Section V.
This paper is summarized in Section VI.
Table \ref{tab1} presents the main parameters in this paper.

\begin{table}[!t]
	\renewcommand\arraystretch{1.5}
	\caption{\label{tab1}Key parameters}
	\begin{center}
		\begin{tabular}{l l}
			\hline
			\hline
			{Symbol} & {Description} \\
			\hline
			${f}_{c}$ & Carrier frequency \\			
			${N}_{s}$ & Number of OFDM symbols \\
			${N}_{c}$ & Number of subcarriers  \\
			${\Delta f}$ & Subcarrier interval of OFDM signal \\
			${\varphi_0}$ & Initial phase of OFDM signal \\
			${T}_{0}$ & Transmission time of OFDM signal \\
			${T}$ & Full duration of an OFDM symbol \\
			${R_n}$ & Distance between target and BS $n$\\
			$v_n$ & Radial velocity of target with respective to BS $n$\\
			${C}$ & Propagation velocity of signal\\
			${U_n}$ & Attenuation of echo signal received by BS $n$\\
			${\theta_n}$ & The direction of BS $n$ with respective to target\\
			${\bf{d}}_{n,m}$ & QAM symbol\\			
			\hline
			\hline
		\end{tabular}
	\end{center}
\end{table}

\section{Scenario And Signal Model}
\subsection{Scenario}
Suppose that there are a total of $W$ scattered BSs in an area, where the coordinates of the $w$th BS is $({x_w},{y_w})$ and all BSs are connected using optical fibers. A target is located within the sensing range of all BSs and moving in a certain direction at a certain velocity. When multiple BSs engage in cooperative sensing, all BSs transmit orthogonal ISAC signals to the approximate area of the target and receive the echo signals of their own transmitted signals for further signal processing. After preliminary processing of ISAC signals, each BS uploads the data of preliminary processed signal to the fusion center through optical fiber and performs sensing information fusion.

\begin{figure}[!ht]
	\centering
	\includegraphics[width=0.48\textwidth]{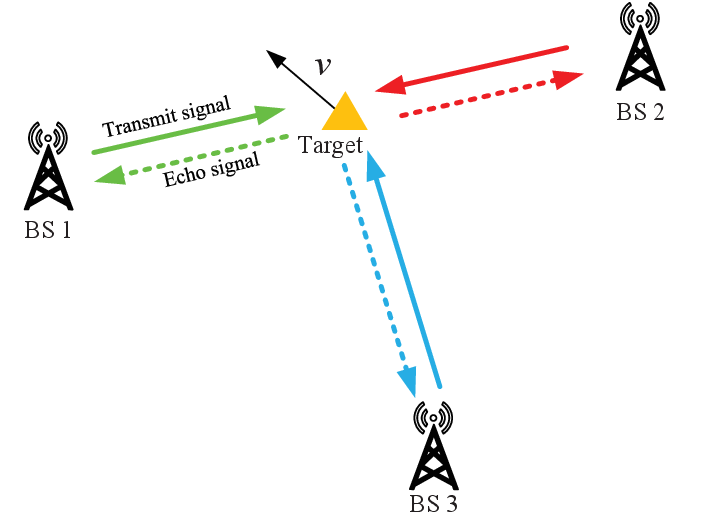}
	\caption{Multiple BSs cooperative sensing.}
	\label{fig2}
\end{figure}

As shown in Fig. \ref{fig2}, 
the solid lines represent the transmitted ISAC signals, 
while the dashed lines represent the echo signals.
The triangle represents the target and 
the black arrow represents the velocity vector
of target.

\subsection{Transmit signal model}
In this paper, OFDM signal is used as ISAC signal, which realizes communication and target sensing. The ISAC frequency band signal transmitted by single-BS is 
\begin{equation}\label{eq1}
\begin{aligned}
	{S_{TX}}(t) = & \sum\limits_{n = 1}^{{N_s}} {\sum\limits_{m = 0}^{{N_c} - 1} {\left[ {{{\rm Re}} \left( {{{\bf{d}}_{n,m}}{e^{j\left( {2\pi ({f_c} + m\Delta f)t + {\varphi _0}} \right)}}} \right)} \right.} } \\
	& \left. { \times {\rm rect} \left( {\frac{{t - {\rm{ }}{T_0} - (n - 1)T}}{T}} \right)} \right],
\end{aligned}
\end{equation}
where $T_0$ is the time instant when the first symbol is sent, 
$N_s$ is the number of OFDM symbols, 
$n$ represents the index of OFDM symbol,
$N_c$ is the number of subcarriers,
$m$ represents the index of subcarrier, 
${\bf{d}}_{n,m}$  is the QAM symbol modulated 
on the $m$-th subcarrier of the $n$-th OFDM symbol, 
$\varphi _0$ is the initial phase of transmitted frequency band signal,
$\Delta f = \frac{1}{T}$ is the interval between subcarriers,
$T$ is the duration of an OFDM symbol. 
Besides, $\rm rect (*)$ is a window function, with expression
\begin{equation}\label{eq2}
	{\rm rect} \left(\frac{{t}}{T}\right) = \left\{ {\begin{array}{*{20}{c}}
			1&{0 < t < T}\\
			0&{{\rm{otherwise}}}
	\end{array}} \right..
\end{equation}

\subsection{Received echo signal model}
The received echo ISAC signal is used in target sensing.
When the distance and radial velocity of the target to the BS are  $R$ and $v$, the received echo signal has expression as follows.
\begin{equation}\label{eq3}
\begin{aligned}
	{S_{RX}}(t) = & U\sum\limits_{n = 1}^{{N_s}} {\sum\limits_{m = 0}^{{N_c} - 1} {\left[ {{\rm{Re}}\left( {{{\bf{d}}_{n,m}}{e^{j\left( {2\pi \left( {{f_c} + m\Delta f} \right)\left( {t - \frac{{2R - 2vt}}{C}} \right) + {\varphi _0}} \right)}}} \right)} \right.} } \;\\
	& \left. { \times {\rm{rect}}\left( {\frac{{t - {T_0} - (n - 1)T - \frac{{2R}}{C}}}{T}} \right)} \right] + Z(t),
\end{aligned}
\end{equation}
where $U$ denotes the radar sensing channel gain \cite{19} 
and $Z(t)$ is the additive white Gaussian noise.

The number of samples of FFT during demodulation is set to $N_c$, and ${{{\bf{d}}'}_{n,m}}$ is the demodulation symbol in the $m$-th subcarrier of the $n$-th OFDM symbol in the received echo signal.
\begin{equation}\label{eq4}
{{{\bf{d}}'}_{n,m}} = U'{{\bf{d}}_{n,m}}{e^{ - j2\pi m\Delta f\frac{{2R}}{C}}}{e^{j2\pi {f_c}\frac{{2vnT}}{C}}} + Z(n,m),
\end{equation}
where the expression of the variable ${U}^{'}$ is
\begin{equation}\label{eq5}
U' = \frac{{U}}{2}N_c{e^{ - j2\pi {f_c}\frac{{2R}}{C}}}{e^{-j2\pi {f_c}\frac{{2vT_s}}{C}}},
\end{equation}
$T_s$ is the duration of an OFDM symbol without cyclic prefix and $Z(n,m)$ is the noise component in the $m$-th subcarrier of the $n$-th OFDM symbol.

\section{Symbol-level Multi-BS Cooperative Sensing}
\begin{figure*}
	\begin{equation}\label{eq6}
		{\bf{B}}_w=U_w \left(
		\begin{matrix}
			1 & e^{j2\pi f_c\frac{2v_wT}{C}} & \cdots  & e^{j2\pi f_c\frac{2v_w\left( N_s-1 \right)T}{C}}  \\
			e^{-j2\pi \Delta f\frac{2R_w}{C}} & e^{-j2\pi \Delta f\frac{2R_w}{C}} e^{j2\pi f_c\frac{2v_wT}{C}} & \cdots  & e^{-j2\pi \Delta f\frac{2R_w}{C}} e^{j2\pi f_c\frac{2v_w\left( N_s-1 \right)T}{C}}  \\
			\vdots  & \vdots  & \ddots  & \vdots   \\
			e^{-j2\pi \left( N_c-1 \right)\Delta f\frac{2R_w}{C}} & e^{-j2\pi \left( N_c-1 \right)\Delta f\frac{2R_w}{C}} e^{j2\pi f_c\frac{2v_wT}{C}} & \cdots  & e^{-j2\pi \left( N_c-1 \right)\Delta f\frac{2R_w}{C}} e^{j2\pi f_c\frac{2v_w\left( N_s-1 \right)T}{C}}  \\
		\end{matrix} \right).
	\end{equation}
\end{figure*}
In multi-BS cooperative sensing, 
the sensing information fusion is essential. 
In symbol-level fusion, 
the demodulation symbols are firstly obtained from the echo signals of multiple BSs,
then the phase features contained in the demodulation symbols are used in the fusion procedure, 
which is different from data-level fusion and signal-level fusion.
In multi-BS cooperative sensing, 
single-BS signal preprocessing is firstly performed. 
Then, multi-BS sensing information fusion is performed 
for the location and velocity estimation of target. 

\subsection{Single-BS Signal Preprocessing}
\label{sec_single}

In the signal preprocessing process of single-BS, 
it is necessary to remove the information unrelated to the sensing information of target
in the received echo signal, 
retaining only the distance and velocity information of target 
and minimizing the loss of information during signal processing as much as possible.

\subsubsection{Communication Information Separation}
In order to mitigate the influence of communication information on sensing, 
the demodulation symbols of the received echo signal are correspondingly 
divided by the modulation symbols of the transmit signal in \cite{19}, 
so that the communication information is eliminated. 
The symbols that remove the communication information are arranged into a matrix ${\bf{B}}_w$ according to the subcarrier order and time order in (\ref{eq6}) at the top of the next page, where $U_w$ is the radar sensing channel gain of BS $w$, $v_w$ is the radial velocity of target with respective to BS $w$, $R_w$ is the distance of target with respective to BS $w$.

\subsubsection{Joint Range and Velocity Estimation}
\label{sec_III_A_2}
Though the phase change of the elements in each row and each column of 
the matrix ${\bf{B}}_w$ reveals the radial velocity and the distance of target to the BS $w$, 
the data is further processed for the following reasons.

\begin{enumerate}
	\item The information of the target's distance in the complex coefficient of 
	the same subcarrier on different symbols is the same, 
	and the information of the target's velocity in the complex coefficient of 
	different subcarriers on the same symbol is the same.
	\item With the matched filtering, 
	the signals of multiple subcarriers are separated from each other. 
	However, since the noise is not orthogonal to the signal on any subcarrier, 
	the power of noise does not decrease after the matched filtering operation, 
	resulting in low SNR for each element in matrix ${\bf{B}}_w$.
	\item The number of elements in matrix ${\bf{B}}_w$ is large. 
	Using the matrix ${\bf{B}}_w$ in sensing information fusion will 
	result in high computational complexity.
\end{enumerate}

Hence, it is necessary to compress the elements of matrix ${\bf{B}}_w$ 
and use the obtained results in subsequent multi-BS sensing information fusion, 
which will be introduced in Section \ref{sec_3a3}.

Because each element in ${\bf{B}}_w$ contains different delay and Doppler frequency shift, the phases of the elements in ${\bf{B}}_w$ need to be adjusted before fusion, and an improved two-dimensional FFT (2D FFT) realizes the approximate estimation of the distance and radial velocity of the target relative to the single-BS.

Assuming that the distance from the target to the single-BS is between $\left[{R}_{\rm{min}},{R}_{\rm{max}} \right]$, 
which is sampled into $K$ evenly distributed sampling points 
denoted by $R'_1={R}_{\rm{min}}, R'_2,R'_3,\cdots,R'_{K-1},R'_K = {R}_{\rm{max}}$,
with $\Delta R=\frac{{{R}_{\rm{max}}}-{{R}_{\rm{min}}}}{K}$ as interval. 
For the sampling point $R'_k, k=1,2,\cdots,K$, 
a distance compensation vector ${\bf{A}}_k$ is 
\begin{equation}\label{eq7}
	{\bf{A}}_k=\left(
	\begin{matrix}
		1 & e^{j2\pi \Delta f\frac{2{R'_k}}{C}} & \cdots  & e^{j2\pi \left( N_c-1 \right)\Delta f\frac{2{R'_k}}{C}}  \\
	\end{matrix} \right).
\end{equation}

Arranging all the distance compensation vectors in the order of distance 
from small to large, 
a distance compensation matrix ${\bf{A}}$ is formed as follows.
\begin{equation}\label{eq8}
	{\bf{A}}=\left( \begin{matrix}
		1 & e^{j2\pi \Delta f\frac{2{R'_1}}{C}} & \cdots  & e^{j2\pi \left( N_c-1 \right)\Delta f\frac{2{R'_1}}{C}}  \\
		1 & e^{j2\pi \Delta f\frac{2{R'_2}}{C}} & \cdots  & e^{j2\pi \left( N_c-1 \right)\Delta f\frac{2{R'_2}}{C}}  \\
		\vdots  & \vdots  & \ddots  & \vdots   \\
		1 & e^{j2\pi \Delta f\frac{2{R'_K}}{C}} & \cdots  & e^{j2\pi \left( N_c-1 \right)\Delta f\frac{2{R'_K}}{C}}  \\
	\end{matrix} \right).
\end{equation}

Since the moving direction of target is unknown, its radial velocity with respective to a BS may be positive or negative. Assuming that the radial velocity is in the range of [$v_{\rm{min}}$, $v_{\rm{max}}$], which is sampled into $P$ evenly distributed sampling points denoted by $v'_1={v}_{\rm{min}}, v'_2,v'_3,\cdots,v'_{P-1},v'_P = {v}_{\rm{max}}$, with $\Delta v=\frac{{{v}_{\rm{max}}}-{{v}_{\rm{min}}}}{P}$ as interval. For the sampling point $v_p,p=1,2,\cdots,P$, a velocity compensation vector ${\bf{C}}_p$ is
\begin{equation}\label{eq9}
	{\bf{C}}_p={\left(
		\begin{matrix}
			1 & e^{-j2\pi f_c\frac{2v'_pT}{C}} & \cdots  & e^{-j2\pi f_c\frac{2v'_p\left( N_s-1 \right)T}{C}}  \\
		\end{matrix} \right)^{T}}.
\end{equation}

Arranging all the velocity compensation vectors in the order of velocity from small to large, a velocity compensation matrix ${\bf{C}}$ is formed as (\ref{eq10}).
\begin{equation}\label{eq10}
	{\bf{C}}=\left(
	\begin{matrix}
		1 & \cdots & 1 \\
		e^{-j2\pi f_c\frac{2v'_1T}{C}} & \cdots  & e^{-j2\pi f_c\frac{2v'_PT}{C}}  \\
		\vdots & \ddots & \vdots   \\
		e^{-j2\pi f_c\frac{2v'_1\left( N_s-1 \right)T}{C}} & \cdots  & e^{-j2\pi f_c\frac{2v'_P\left( N_s-1 \right)T}{C}}  \\
	\end{matrix} \right).
\end{equation}

\begin{figure*}[!ht]
	\begin{equation}\label{eq11}
		\begin{aligned}
			{\bf{D}}_w  & = {\bf{AB}}_w{\bf{C}}\\
			&	= U_w \left( {\begin{array}{*{20}{c}}
					\begin{array}{l}
						\sum\limits_{m = 0}^{N_c  - 1} {e^{j2\pi m\Delta f\frac{{2\left( {R_1  -  R_w } \right)}}{C}} }  \times \\
						\sum\limits_{n = 0}^{N_s-1} {e^{j2\pi f_c \frac{{2(v_w  - v_1 )nT}}{C}} }
					\end{array}&\begin{array}{l}
				    	\sum\limits_{m = 0}^{N_c  - 1} {e^{j2\pi m\Delta f\frac{{2\left( {R_1  - R_w } \right)}}{C}} }  \times \\
					    \sum\limits_{n = 0}^{N_s-1} {e^{j2\pi f_c \frac{{2(v_w  - v_2 )nT}}{C}} }
			    	\end{array}& \cdots &\begin{array}{l}
						\sum\limits_{m = 0}^{N_c  - 1} {e^{j2\pi m\Delta f\frac{{2\left( {R_1  - R_w } \right)}}{C}} }  \times \\
						\sum\limits_{n = 0}^{N_s-1} {e^{j2\pi f_c \frac{{2(v_w  - v_P )nT}}{C}} }
					\end{array}\\
					\vdots & \vdots & \ddots & \vdots \\
					\begin{array}{l}
						\sum\limits_{m = 0}^{N_c  - 1} {e^{j2\pi m\Delta f\frac{{2\left( {R_K  - R_w } \right)}}{C}} }  \times \\
						\sum\limits_{n = 0}^{N_s-1} {e^{j2\pi f_c \frac{{2(v_w  - v_1 )nT}}{C}} }
					\end{array}&\begin{array}{l}
					    \sum\limits_{m = 0}^{N_c  - 1} {e^{j2\pi m\Delta f\frac{{2\left( {R_K  - R_w } \right)}}{C}} }  \times \\
					    \sum\limits_{n = 0}^{N_s-1} {e^{j2\pi f_c \frac{{2(v_w  - v_2 )nT}}{C}} }
				    \end{array}& \cdots &\begin{array}{l}
						\sum\limits_{m = 0}^{N_c  - 1} {e^{j2\pi m\Delta f\frac{{2\left( {R_K  - R_w } \right)}}{C}} }  \times \\
						\sum\limits_{n = 0}^{N_s-1} {e^{j2\pi f_c \frac{{2(v_w  - v_P )nT}}{C}} }
					\end{array}
			\end{array}} \right)
		\end{aligned}.
	\end{equation}
\end{figure*}

\begin{figure*}[!ht]
	\centering
	\subfigure[The searching area of multi-BS cooperative sensing.]{\includegraphics[width=0.4\textwidth]{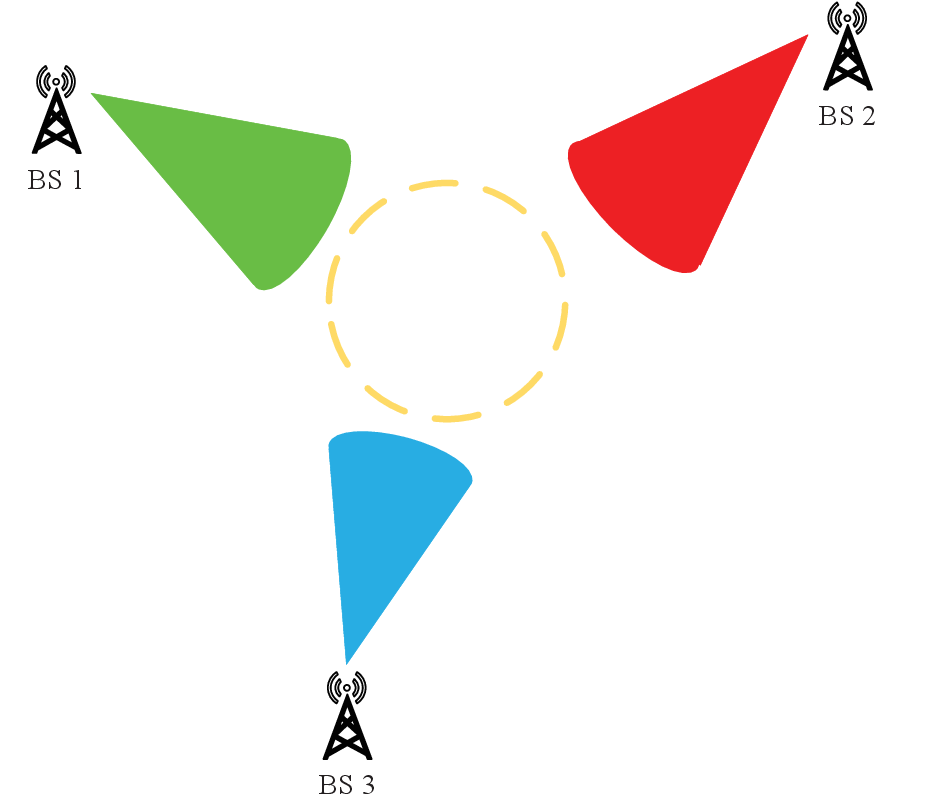}\label{fig3a}}
	\quad
	\subfigure[The area around the roughly estimated target's location is divided into lattice points.]{\includegraphics[width=0.5\textwidth]{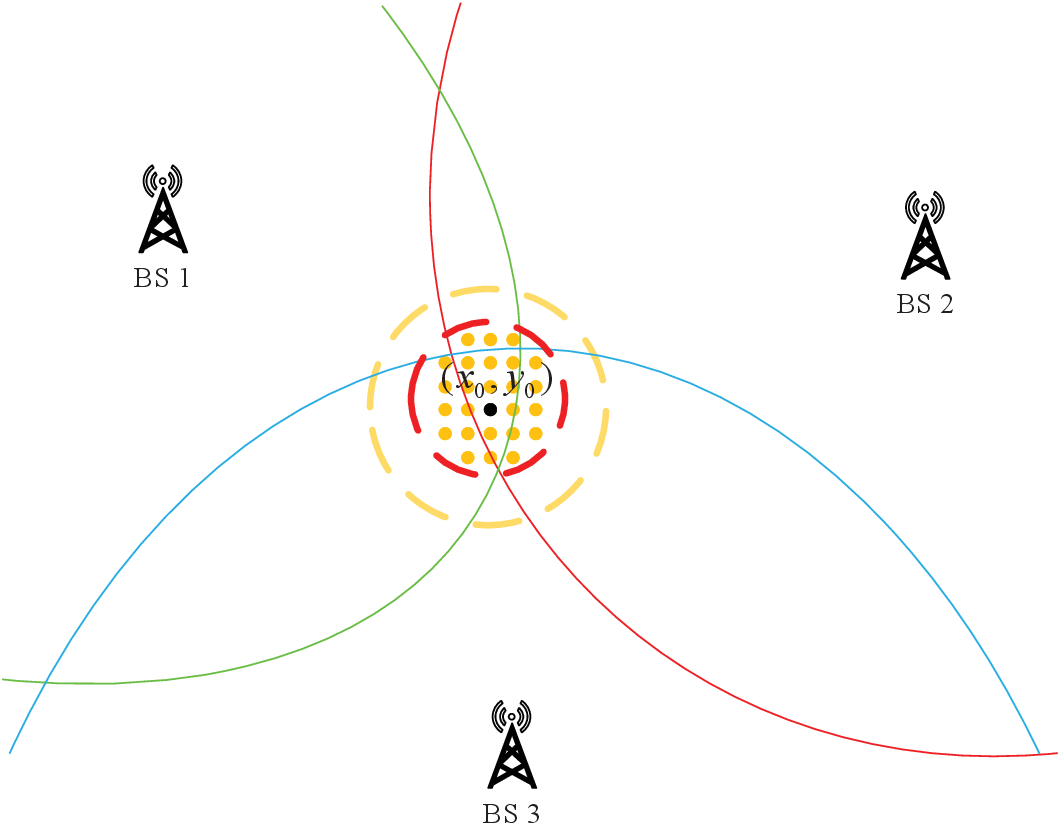}\label{fig3b}}
	\caption{Estimation of the location of target.}
	\label{fig3ab}
\end{figure*}

The matrices $\bf{A}$, ${\bf{B}}_w$, and $\bf{C}$ are multiplied, yielding an estimation matrix ${\bf{D}}_w$,
as shown in (\ref{eq11}) at the top of the next page. 
In matrix ${\bf{D}}_w$, the indices of the element with largest modulus corresponds to the estimated distance and velocity of target. According to the indices $\left(x_{\rm{index}},y_{\rm{index}}\right)$ of this element in the matrix ${\bf{D}}_w$, the corresponding distance $R_{w,\rm{test}}=R'_{{x}_{\rm{index}}}$ and radial velocity $v_{w,\rm{test}}=v'_{{y}_{\rm{index}}}$ of target with respective to BS $w$ are estimated.

\subsubsection{Matrix Compression}\label{sec_3a3}

After obtaining the estimated distance and radial velocity of the target relative to the BS, the elements of matrix ${\bf{B}}_w$ can be compressed to obtain a symbol vector ${\bf{E}}_{w}$ that preserves the distance information of target and a symbol vector ${\bf{F}}_{w}$ that preserves the radial velocity information of target.

The velocity compensation vector ${\bf{C}}_{\rm{test}}$ corresponding 
to the estimated radial velocity $v_{w,\rm{test}}$ is 
\begin{equation}\label{eq12}
	{\bf{C}}_{\rm{test}}=\left(
		\begin{matrix}
			1 & e^{-j2\pi f_c \frac{2v_{w,test}T}{C}} & \cdots  & e^{-j2\pi f_c\frac{2v_{w,test}\left( N_s-1 \right)T}{C}}  \\
		\end{matrix} \right)^{\rm{T}}.
\end{equation}

The matrices ${\bf{B}}_w$ and ${\bf{C}}_{\rm{test}}$ are multiplied, 
yielding the vector ${\bf{E}}_w$ as follows.

\begin{equation}\label{eq13}
	\begin{aligned}
	{\bf{E}}_w  & = {\bf{B}}_w {\bf{C}}_{\rm{test}} \\
	&  = U_{w} \sum\limits_{n = 1}^{N_s} {e^{{j}2\pi f_c \frac{{2(v_w - v_{w,test} )nT}}{C}}} \\ & \quad \times \left( \begin{matrix}
		1 & e^{-j2\pi \Delta f\frac{2R_w}{C}} &	\cdots &	e^{-j2\pi (N_c-1)\Delta f\frac{2R_w}{C}}   \\
	\end{matrix} \right)^{\rm{T}}.
   	\end{aligned}
\end{equation}

As shown in (\ref{eq13}), the phases change of the elements in ${\bf{E}}_w$ is only related to the distance of target. The distance compensation vector ${\bf{A}}_{\rm{test}}$ corresponding to the estimated distance $R_{n,\rm{test}}$ is 
\begin{equation}\label{eq14}
	{\bf{A}}_{\rm{test}}=\left(
	\begin{matrix}
		1 & e^{j2\pi \Delta f\frac{2R_{w,test}}{C}} & \cdots  & e^{j2\pi \left(N_c-1 \right)\Delta f\frac{2R_{w,test}}{C}}  \\
	\end{matrix} \right).
\end{equation}	
	
The matrices ${\bf{A}}_{\rm{test}}$ and ${\bf{B}}_w$ are multiplied, yielding the vector ${\bf{F}}_w$ as follows.

\begin{equation}\label{eq15}
	\begin{aligned}
	{\bf{F}}_w  & = {\bf{A}}_{\rm{test}} {\bf{B}}_w \\
	 &  = U_{w}\sum\limits_{m = 0}^{N_c  - 1} {e^{{j}2\pi m\Delta f\frac{{2(R_{w,test}  - R_w)}}{C}} } \\ &  	\quad \times \left( 
	 \begin{matrix}
		1 & e^{j2\pi f_c\frac{2v_wT}{C}} & \cdots & e^{j2\pi f_c\frac{2v_w\left(N_s-1 \right)T}{C}}   \\
	\end{matrix} \right).
	\end{aligned}
\end{equation}

As shown in (\ref{eq15}), the phase change of the elements in ${\bf{F}}_w$ is only related to the velocity of target. Due to the fact that each element of vector ${\bf{E}}_w$ carries the distance information of the target, it is named the distance feature vector. Similarly, vector ${\bf{F}}_w$ is named the velocity feature vector.

\subsection{Multi-BS Sensing Information Fusion}
\label{sec_multi}

When a BS has preprocessed the received echo signal, 
the BS sends four parameters to the fusion center. 
For example, the parameters of BS $w$ are the estimated distance $R_{w,\rm{test}}$ 
and estimated radial velocity $v_{w,\rm{test}}$ of the target, 
the vectors ${\bf{E}}_w$ and ${\bf{F}}_w$ carrying the phase change characteristics. 
$R_{w,\rm{test}}$ and $v_{w,\rm{test}}$ 
are used to determine the approximate range of the location and the velocity of target, 
while the vectors ${\bf{E}}_w$ and ${\bf{F}}_w$ from multiple BSs
will be fused to achieve accurate estimation of 
the location and velocity of target.

It is noted that the localization of target is firstly performed 
to obtain the direction of target relative to each BS.
Then, multi-BS cooperative velocity estimation can be performed.

\subsubsection{Multi-BS Cooperative Location Estimation}\label{sec_III_B_2}

The multi-BS cooperative location estimation of target consists of two steps,
namely rough estimation of target's location and accurate estimation of target's location.

\textbf{Step 1: Rough estimation of target's location}

When multiple BSs perform cooperative sensing, the searching area should be determined, as shown in the dashed circle in Fig. \ref{fig3a}. According to the estimated distance relative to all BSs, the approximate location of target can be derived. Take BS $Q$ and BS $S$ for an example, assuming that the coordinates of these two BSs are $\left(x_Q,y_Q\right)$ and $\left(x_S,y_S\right)$, the estimated distances between these two BSs and target are $R_{Q,\rm{test}}$ and $R_{S,\rm{test}}$. Then, the coordinates of the location of the target $(x_{\rm{csp}\_QS}, y_{\rm{csp}\_QS})$ are derived in (\ref{eq16}).

\begin{equation}\label{eq16}
	\left\{\begin{array}{l}
		x_{\rm{csp}\_QS}  = \frac{{x_Q  + x_S }}{2} + 
		\frac{{R_{Q,\rm{test}}^2  - R_{S,\rm{test}}^2 }}{{2d_{Q - S}^2 }}\left( {x_S  - x_Q } \right)\\
		\quad \quad \pm \frac{{ y_S  -  y_Q }}{2}\sqrt {2\frac{{ R_{Q,\rm{test}}^2  + R_{S,\rm{test}}^2 }}{{d_{Q - S}^2 }} - \frac{{ {\left( {R_{Q,\rm{test}}^2  - R_{S,\rm{test}}^2 } \right)}^2 }}{{ d_{Q - S}^4 }} - 1},\\
		y_{\rm{csp}\_QS}  = \frac{{ y_Q  +  y_S }}{2} + \frac{{ R_{Q,\rm{test}}^2  -  R_{S,\rm{test}}^2 }}{{2 d_{Q - S}^2 }}\left( { y_S  -  y_Q } \right)\\ \quad \quad \pm \frac{{ x_Q  -  x_S }}{2}\sqrt {2\frac{{ R_{Q,\rm{test}}^2  +  R_{S,\rm{test}}^2 }}{{ d_{Q - S}^2 }} - \frac{{ {\left( { R_{Q,\rm{test}}^2  - R_{S,\rm{test}}^2 } \right)}^2 }}{{ d_{Q - S}^4 }} - 1}.
	\end{array}\right.
\end{equation}

It is noted that (\ref{eq16}) generates two coordinates of target since there are no sufficient BSs. However, the false result could be eliminated by calculating whether the coordinates are within the sensing area. If there are more than two BSs, the approximate location of target $\left(x_0 ,y_0\right)$ shown in Fig. \ref{fig3b} can be estimated by other methods, such as least square method. 
When determining the location of target, 
several lattice points are divided around $\left(x_0 ,y_0\right)$, 
as illustrated in Fig. \ref{fig3b}.
In the next step, the weights of all lattice points will be calculated. 
Then, the lattice point with the largest weight is 
the estimated location of target.

\textbf{Step 2: Accurate estimation of target's location}

\begin{figure}[!ht]
	\centering
	\includegraphics[width=0.5\textwidth]{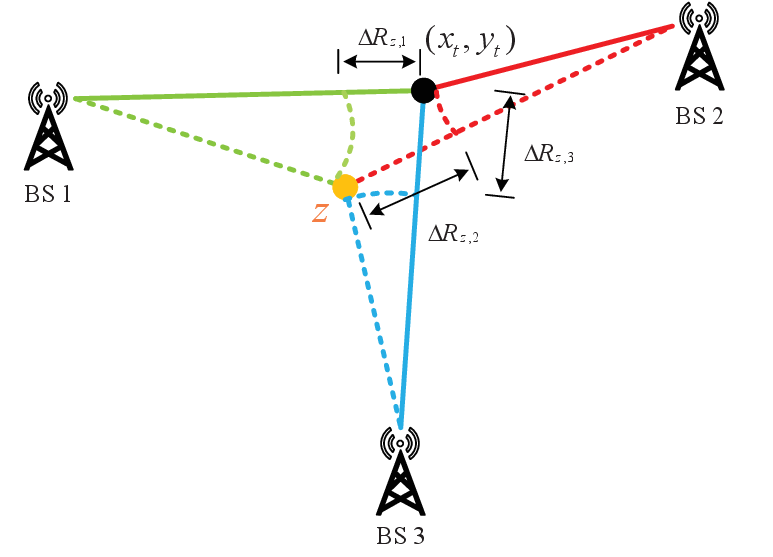}
	\caption{The difference between the lattice point and the real location of target.}
	\label{fig4}
\end{figure}

In order to improve the accuracy of the estimation of target's location, 
the difference of the weights between the lattice point corresponding 
to the target's location and other surrounding lattice points is amplified.
For the lattice point $z$, 
the difference between its distance to BS $w$ and the distance from the target to BS $w$ 
is denoted by $\Delta R_{z,w}$, 
as shown in Fig. \ref{fig4}.

The distance difference between lattice point $z$ 
and the target relative to each BS can be accumulated. 
Then, the total distance difference is 
\begin{equation}\label{eq17}
	\begin{aligned}
		\Delta R_{z,\rm{sum}} = \sum\limits_{w = 1}^{W}{\Delta R_{z,w}}
	\end{aligned}.
\end{equation}

In this way, the difference between the lattice point $z$ 
and the location of target is amplified. 
In order to accumulate the distance difference, 
IDFT is firstly performed on distance feature vectors ${\bf{E}}_1,{\bf{E}}_2,\cdots,{\bf{E}}_W$ 
according to the distance between the lattice point $z$ and each BS, 
so that the phase of the elements of distance feature vectors 
is no longer related to the distances between the target and all the BSs, 
but related to the distance difference values $\Delta R_{z,1},\Delta R_{z,2},\cdots,\Delta R_{z,W}$. 
The distance between the lattice point $z$ and BS $w$ is denoted by $R_{z,w}$. 
Then, the distance feature vector after IDFT is
\begin{equation}\label{eq18}
	\begin{aligned}
		\begin{array}{l}
			\mathop {\bf{E}}_{z,w} = \mathop U_w \sum\limits_{n = 1}^{\mathop N_s } {\mathop e\nolimits^{j2\pi \mathop f_c \frac{{2(\mathop v_w  - \mathop v_{w,test} )nT}}{C}} } \\
			\;\;\quad \;{\kern 1pt} \,\quad  \times \left( {\begin{array}{*{20}{c}}
					1&{\mathop e\nolimits^{j2\pi \Delta f\frac{{2\left( {\mathop R_{z,w}  - \mathop R_w } \right)}}{C}} }
			\end{array}} \right.\\
			\mathop {\quad \quad \quad \quad \;\left. {\begin{array}{*{20}{c}}
						\cdots &{\mathop e\nolimits^{j2\pi \left( {\mathop N_c  - 1} \right)\Delta f\frac{{2\left( {\mathop R_{z,w}  - \mathop R_w } \right)}}{C}} }
				\end{array}} \right)}\nolimits^{\rm{T}} \\
			\;\;\quad \, = \mathop U_w^' \left( {\begin{array}{*{20}{c}}
					1&{\mathop e\nolimits^{j2\pi \Delta f\frac{{2\mathop {\Delta R}_{z,w} }}{C}} }
			\end{array}} \right.\\
			\mathop {\quad \quad \quad \quad \;\left. {\begin{array}{*{20}{c}}
						\cdots &{\mathop e\nolimits^{j2\pi \left( {\mathop N_c  - 1} \right)\Delta f\frac{{2\mathop {\Delta R}_{z,w} }}{C}} }
				\end{array}} \right)}\nolimits^{\rm{T}}. 
		\end{array}
	\end{aligned}
\end{equation}

Then, the Hadamard product of all vectors transformed by IDFT is as follows.
\begin{equation}\label{eq19}
	\begin{aligned}
		\begin{array}{l}
			{\bf{E}}_z   = \prod\limits_{w = 1}^W {U_w^' } \left( {\begin{array}{*{20}{c}}
					1&{e^{j4\pi \Delta f\sum\limits_{w = 1}^{W}{\Delta R_{z,w}}/{C}} }
			\end{array}} \right.\\
			\quad \quad \quad \quad \quad \quad \left. {\begin{array}{*{20}{c}}
					\cdots &{e^{j4\pi (N_c-1) \Delta f\sum\limits_{w = 1}^{W}{\Delta R_{z,w}}/{C}} }
			\end{array}} \right)^{\rm{T}}.
		\end{array}
	\end{aligned}
\end{equation}

The phase of each element of the vector ${\bf{E}}_z$ in (\ref{eq19}) 
is related to the total distance difference.
Besides, the distance difference may be positive or negative. 
In (\ref{eq19}), if the signs of the distance difference values
are opposite, the absolute value of the accumulation of the distance difference values 
is reduced.
It is found that the real parts of all elements in
the vectors ${\bf{E}}_{z,1},{\bf{E}}_{z,2},...,{\bf{E}}_{z,W}$ 
contain cosine function and the cosine function has the following properties.
\begin{equation}\label{eq20}
	\left\{ {\begin{array}{*{20}{c}}
			{\cos \left( x \right) = \cos \left( { - x} \right)}\\
			\left| {\cos \left( x \right)} \right| \le \cos \left( 0 \right)\quad \\
			\mathop {\lim }\limits_{x \to 0} \left| {\cos \left( x \right)} \right| = \left| x \right|
	\end{array}} \right.,
\end{equation}
where $x$ is a real number. 
Due to the properties of cosine functions in (\ref{eq20}), 
replacing the original element in the vectors ${\bf{E}}_{z,1},{\bf{E}}_{z,2},\cdots,{\bf{E}}_{z,W}$
with its real part
can avoid the mutual cancellation between the distance difference values. 

However, the phase of the complex coefficients 
$U'_{1},U'_{2},\cdots,U'_{W}$ in the vectors ${\bf{E}}_{z,1},{\bf{E}}_{z,2},\cdots,{\bf{E}}_{z,W}$ is unknown, 
so that it is impossible to separate the cosine function value 
of each element in the vectors. 
However, this operation can be realized when adjusting the phase of 
the complex coefficients $U'_{1},U'_{2},\cdots,U'_{W}$ to 0. 

Therefore, a vector reconstruction method is proposed to eliminate
the phase of the complex coefficients $U'_{1},U'_{2},\cdots,U'_{W}$. 
The vectors ${\bf{E}}_1,{\bf{E}}_2,\cdots,{\bf{E}}_W$ are reconstructed 
instead of the vectors after IDFT. 
Since the complex coefficient $U'_{1},U'_{2},...,U'_{W}$ are the common factors in the vectors ${\bf{E}}_1,{\bf{E}}_2,...,{\bf{E}}_W$, 
conjugate multiplication of any two different elements in the same vector 
can eliminate the phase of the complex coefficient while preserving the distance information. 
Taking vector ${\bf{E}}_w$ for example, 
if ${\bf{E}}_w(a)$ and ${\bf{E}}_w(a+k)$ are conjugate multiplied, the result is 
\begin{equation}\label{eq21}
	\begin{aligned}
		&\quad {\bf{E}}_w(a) {{\bf{E}}^*_w(a+k)}\\  & = U'_w e^{{-j}2\pi (a-1)\Delta f\frac{2R_w}{C}} {\left( { U'_w } \right)}^* e^{{j}2\pi (a+k-1)\Delta f\frac{2R_w}{C}}  \\
		&  =  {\left| { U'_w } \right|}^2 e^{j 2\pi k\Delta f\frac{2R_w}{C}},
	\end{aligned}
\end{equation}
where $*$ represents conjugation. 
In vector ${\bf{E}}_w$, 
there are ${N_c-k}$ pairs of elements with the difference of index being $k$. 
The signal components with conjugate multiplication of 
different pairs of elements are the same, 
while the noise components are different. 
The results of conjugate multiplication of different pairs of elements are selected 
as the elements of a new vector ${\bf{G}}_w$.
The SNR of the element in ${\bf{G}}_w$ is higher than that of the element in ${\bf{E}}_w$. 
In this way, the vector ${\bf{G}}_w$ corresponding to the vector ${\bf{E}}_w$ is constructed, 
where the phases of the elements  
are only related to the distance of target. 
The expression for the $k$-th element of ${\bf{G}}_w$ is 
\begin{equation}\label{eq22}
	\begin{aligned}
		{\bf{G}}_w(k) & =\frac{1}{N_c-k} \sum\limits_{a=1}^{N_c-k}{{\bf{E}}_w(a) } {{\bf{E}}_w^*}(a+k) \\
		& = {\left| { U'_w } \right|}^2 e^{j2\pi k\Delta f\frac{2R_w}{C}},
	\end{aligned}
\end{equation}
where $k=1,2,...,N_c-1$. 
The expression for vector ${\bf{G}}_w$ is 
\begin{equation}\label{eq23}
	\begin{aligned}
		\begin{array}{l}
			{\bf{G}}_{w} = {\left| { U'_w } \right|}^2 \left( {\begin{array}{*{20}{c}}
					{e^{j2\pi \Delta f\frac{{2{R}_w }}{C}} }&{e^{j2\pi 2 \Delta f\frac{{2{R}_w }}{C}} }
			\end{array}} \right.\\
			\mathop {\quad \quad \quad \quad \quad \quad\;\left. {\begin{array}{*{20}{c}}
						\cdots &{e^{j2\pi \left( {N_c  - 1} \right)\Delta f\frac{{2{R}_{w} }}{C}} }
				\end{array}} \right)}.
		\end{array}
	\end{aligned}
\end{equation}

Repeating (\ref{eq21}) $\sim$ (\ref{eq23}),
new vectors corresponding to the vectors ${\bf{E}}_1,{\bf{E}}_2,\cdots,{\bf{E}}_W$
are constructed. 
With the elimination of all phases of the complex coefficients $U'_{1},U'_{2},\cdots,U'_{W}$, 
DFT is performed on vectors ${\bf{G}}_{1},{\bf{G}}_{2},...{\bf{G}}_{W}$ 
according to the distance between the lattice point $z$ and each BS. 
The expression of each vector after DFT is similar to (\ref{eq18}), which is 
\begin{equation}\label{eq24}
	\begin{aligned}
		\begin{array}{l}
			{\bf{G}}_{z,w} = {\left| { U'_w } \right|}^2 \left( {\begin{array}{*{20}{c}}
					{e^{-j2\pi \Delta f\frac{{2{\Delta R}_{z,w} }}{C}} }&{e^{-j2\pi 2 \Delta f\frac{{2{\Delta R}_{z,w} }}{C}} } 
			\end{array}} \right.\\
			\mathop {\quad \quad \quad \quad \quad \quad\;\left. {\begin{array}{*{20}{c}}
						\cdots&{e^{-j2\pi \left( {N_c  - 1} \right)\Delta f\frac{{2{\Delta R}_{z,w} }}{C}} }
				\end{array}} \right)},
		\end{array}
	\end{aligned}
\end{equation}
where $\Delta R_{z,w}=R_{z,w}-R_w$. 
Afterwards, the values of cosine function in the real parts of all elements in ${\bf{G}}_{z,1},{\bf{G}}_{z,2},\cdots,{\bf{G}}_{z,W}$ are taken to replace the original values. 
The new vectors are denoted by ${\bf{G}}'_{z,1},{\bf{G}}'_{z,2},\cdots,{\bf{G}}'_{z,W}$, 
with expressions as follows.
\begin{equation}\label{eq25}
	\begin{aligned}
		\begin{array}{l}
			{\bf{G}}'_{z,w} = {\left| { U'_w } \right|}^2 \left( {\begin{array}{*{20}{c}}
					{\cos(2\pi \Delta f\frac{{2{\Delta R}_{z,w} }}{C}) }&{\cos(2\pi 2 \Delta f\frac{{2{\Delta R}_{z,w} }}{C}) } 
			\end{array}} \right.\\
			\mathop {\quad \quad \quad \quad \quad \quad\;\left. {\begin{array}{*{20}{c}}
						\cdots&{\cos(2\pi \left( {N_c  - 1} \right)\Delta f\frac{{2{\Delta R}_{z,w} }}{C}) }
				\end{array}} \right)}.
		\end{array}
	\end{aligned}
\end{equation}

Finally, the inner product of all vectors ${\bf{G}}'_{z,1},{\bf{G}}'_{z,2},...{\bf{G}}'_{z,W}$
is calculated, which is adopted 
as the weight of the lattice point $z$. 
The weight value of lattice point $z$ is denoted by ${\bf{H}}\left( z \right)$, 
with expression as follows.
\begin{equation}\label{eq26}
	\begin{aligned}
		{\bf{H}}\left( z \right) & = \sum\limits_{k = 1}^{N_c - 1}{\prod\limits_{w = 1}^{W} \left[{\bf{G}}'_{z,w}(k)\right] }  \\
		&	= \sum\limits_{k = 1}^{N_c - 1} {\prod\limits_{w = 1}^{W} \left[{{\left| { U'_w } \right|}^2 \cos \left( {2\pi k\Delta f\frac{{2 {\Delta R}_{z,w} }}{C}} \right)}\right] }.
	\end{aligned}
\end{equation}

Repeating (\ref{eq24}) $\sim$ (\ref{eq26}), the weights of all lattice points
are calculated.
Then, the location of the lattice point with the maximum weight value 
is adopted as the estimated location of target.

\begin{figure*}[!ht]
	\centering
	\subfigure[The radial velocity of target with respective to each BS.]{\includegraphics[width=0.48\textwidth]{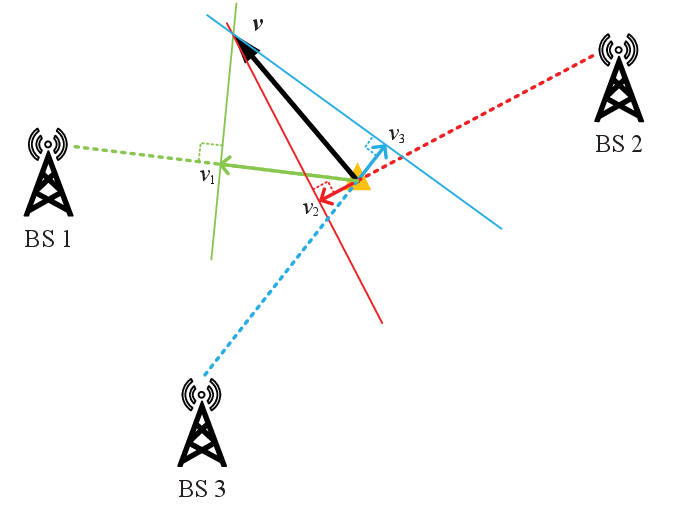}\label{fig5a}}
	\quad
	\subfigure[The area around the roughly estimated target's velocity vector is divided into lattice points.]{\includegraphics[width=0.48\textwidth]{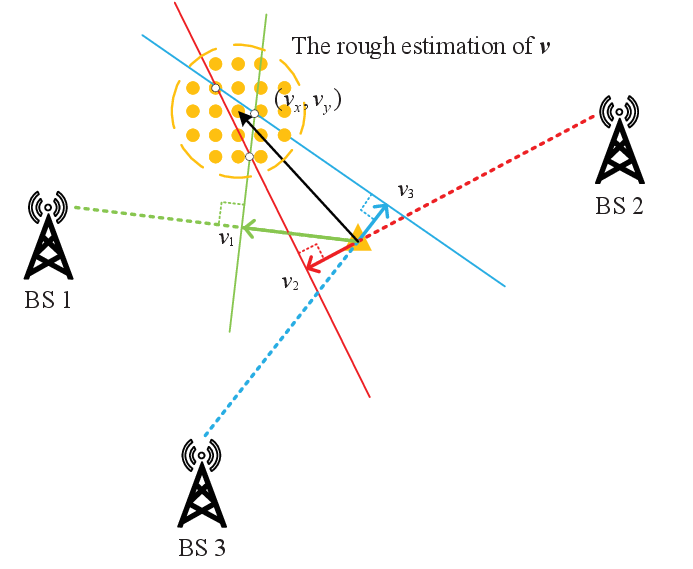}\label{fig5b}}
	\caption{Estimation of the velocity of target.}
	\label{fig5ab}
\end{figure*}

\subsubsection{Multi-BS Cooperative Velocity Estimation}

Similar to Section \ref{sec_III_B_2}, multi-BS cooperative velocity estimation consists of two steps, namely rough and accurate estimation of the velocity of target.

\textbf{Step 1: Rough estimation of target's velocity}

As shown in Fig. \ref{fig5a} in the next page, 
the vectors ${\bf{v}}_1$, ${\bf{v}}_2$ and ${\bf{v}}_3$ represent 
the radial velocities of target with respective to three BSs
and the vector ${\bf{v}}$ represents the velocity vector of target. 
With $(0,0)$ as the coordinate of origin,
the vertical line passing through the end point of 
each radial velocity vector will intersect at one point, 
whose coordinate is the velocity vector of target.

Since the location of target is obtained in Section \ref{sec_III_B_2}, 
the direction of target 
with respective to each BS is determined. Denoting the direction of BS $w$ 
with respective to the estimated location of target as $\theta_w$, 
the radial velocity vector of target with respective to BS $w$ is
\begin{equation}\label{eq27}
	{\bf{v}}_{w,\rm{test}}=\left( v_{w,\rm{test}} \cos (\theta _{w}),v_{w,\rm{test}} \sin (\theta_{w}) \right).
\end{equation}

Taking BS $Q$ and BS $S$ for an example, the equations of the vertical lines 
respective to the radial velocity vectors ${\bf{v}}_Q$ and ${\bf{v}}_S$ are 
\begin{equation}\label{eq28}
	\begin{aligned}
		& y_1=\frac{v_{Q,\rm{test}}}{\sin (\theta_{Q})}-\frac{\cos (\theta_{Q})}{\sin (\theta_{Q})}x, \\
		& y_2=\frac{{v_{S,\rm{test}}}}{\sin (\theta_{S})}-\frac{\cos (\theta_{S})}{\sin (\theta_{S})}x.
	\end{aligned}
\end{equation}

The coordinate of the intersection of the two vertical lines is
the coordinate of the roughly estimation of ${\bf{v}}$ represented by the vector $({v_{x,QS}},{v_{y,QS}})$, with
\begin{equation}\label{eq29}
	\begin{aligned}
		&	{v_{x,QS}} = \frac{{{{v_{Q,\rm{test}}}} \sin (\theta_S) - {{v_{S,\rm{test}}}} \sin (\theta_Q)}}{{\cos (\theta_Q)\sin (\theta_S) - \cos (\theta_S)\sin (\theta_Q)}},\\
		&	{v_{y,QS}} = \frac{{{{v_{S,\rm{test}}}} \cos (\theta_Q) - {{v_{Q,\rm{test}}}} \cos (\theta_S)}}{{\cos ( \theta_Q)\sin (\theta_S) - \cos (\theta_S)\sin (\theta_Q)}}.
	\end{aligned}
\end{equation}

When there are more than two BSs, the coordinates of the velocity vector of the target can be estimated by the methods such as 
least square method. As illustrated in Fig. \ref{fig5b}, 
taking the rough estimation of the velocity vector $({v_x},{v_y})$ as the center, the area around the center is divided into lattice points.

\textbf{Step 2: Accurate estimation of target's velocity}

According to the target location estimated in Section \ref{sec_III_B_2}, the unit vectors denoting the direction from target to all BSs take the form of
\begin{equation}\label{eq30}
	{{\bf{a}}_{t,w}} = \frac{{({x_w} - {x_{t,e}},{y_w} - {y_{t,e}})}}{{\sqrt {{{\left( {{x_w} - {x_t}} \right)}^2} + {{\left( {{y_w} - {y_t}} \right)}^2}} }},
\end{equation}
where $(x_{t,e}, y_{t,e})$ is the estimated location of target and $(x_w, y_w)$ is the location of BS $w$. 
When calculating the weight value of the lattice point $q$, 
the velocity vector coorsponding to lattice point $q$ is ${\bf{v}}_{q}$, 
and the radial velocity of the lattice point $q$ with respective to BS $w$ is 
\begin{equation}\label{eq31}
	v_{q,w}  = {\bf{v}}_{q} \cdot {\bf{a}}_{t,w},
\end{equation}
where ``$\cdot$'' denotes the inner product of two vectors.

Similar to (\ref{eq22}), 
the new vectors ${\bf{I}}_1,{\bf{I}}_2,\cdots,{\bf{I}}_W$ corresponding to the vectors ${\bf{F}}_1,{\bf{F}}_2,\cdots,{\bf{F}}_W$ are constructed, 
where the phases of elements are only related to the radial distance of the target. 
The expression for the $k$-th element of ${\bf{I}}_w$ is 
\begin{equation}\label{eq32}
	\begin{aligned}
		{\bf{I}}_w (k) & =\frac{1}{N_s-k} \sum\limits_{a = 1}^{N_s  - k} {{\bf{F}}_w(a){\bf{F}}_w^*(a + k)} \\
		&\begin{array}{l}
			= \frac{1}{{N_s-k}}\sum\limits_{a = 1}^{N_s-k} {\left[ U''_w e^{j2\pi f_c\frac{2v_w\left( a-1 \right)T}{C}} \right.} \\
			\quad \quad \quad \quad \quad \quad \left. {\left( U''_w  \right)}^* e^{-j2\pi f_c\frac{2v_w\left( {a+k}-1 \right)T}{C}} \right]
		\end{array}\\
		& = {\left| U''_w \right|}^2 e^{-j2\pi f_c\frac{2v_w k T}{C}},
	\end{aligned}
\end{equation}
where
\begin{equation}\label{eq33}
	U''_w={U_{w}}\sum\limits_{m = 0}^{N_c  - 1} { e^{j2\pi m\Delta f\frac{{2( R_{w,test}  - R_w)}}{C}} },
\end{equation}
and the expression for vector ${\bf{I}}_w$ is 
\begin{equation}\label{eq34}
	\begin{aligned}
		\begin{array}{l}
			{\bf{I}}_{w} = {\left| { U''_w } \right|}^2 \left( {\begin{array}{*{20}{c}}
					{e^{-j2\pi f_c\frac{2v_w T}{C}} }&{e^{-j2\pi f_c\frac{4v_w T}{C}} }
			\end{array}} \right.\\
			\mathop {\quad \quad \quad \quad \quad \quad\;\left. {\begin{array}{*{20}{c}}
						\cdots &{e^{-j2\pi f_c\frac{2v_w (N_s-1) T}{C}} }
				\end{array}} \right)}.
		\end{array}
	\end{aligned}
\end{equation}
Repeating (\ref{eq32}) and (\ref{eq34}), 
the new vectors corresponding to the vectors ${\bf{F}}_1,{\bf{F}}_2,\cdots,{\bf{F}}_W$ are constructed.

With the elimination of the phases of complex coefficients $U''_{1},U''_{2},\cdots,U''_{W}$, 
IDFT is performed on the vectors ${\bf{I}}_{1},{\bf{I}}_{2},\cdots,{\bf{I}}_{W}$ 
according to the radial velocity of the lattice point $q$ with respective to each BS. 
The expression of each vector after IDFT is
\begin{equation}\label{eq35}
	\begin{aligned}
		\begin{array}{l}
			{\bf{I}}_{q,w} = {\left| { U''_w } \right|}^2 \left( {\begin{array}{*{20}{c}}
					{e^{j2\pi f_c\frac{2(v_{q,w}-v_w) T}{C}} }&{e^{j2\pi f_c\frac{4(v_{q,w}-v_w) T}{C}} }
			\end{array}} \right.\\
			\mathop {\quad \quad \quad \quad \quad \quad\;\left. {\begin{array}{*{20}{c}}
						\cdots &{e^{j2\pi f_c\frac{2(v_{q,w}-v_w) (N_s-1) T}{C}} }
				\end{array}} \right)}\\
			\;\;\quad \,= {\left| { U''_w } \right|}^2 \left( {\begin{array}{*{20}{c}}
					{e^{j2\pi f_c\frac{2\Delta v_{q,w} T}{C}} }&{e^{j2\pi f_c\frac{4\Delta v_{q,w} T}{C}} }
			\end{array}} \right.\\
			\mathop {\quad \quad \quad \quad \quad \quad\;\left. {\begin{array}{*{20}{c}}
						\cdots &{e^{j2\pi f_c\frac{2\Delta v_{q,w} (N_s-1) T}{C}} }
				\end{array}} \right)},
		\end{array}
	\end{aligned}
\end{equation}
where $\Delta v_{q,w}=v_{q,w}-v_w$. 
Then, the values of cosine function in the real parts of the elements in ${\bf{I}}_{q,1},{\bf{I}}_{q,2},...{\bf{I}}_{q,W}$ are taken to replace the original values. 
The new vectors are denoted by ${\bf{I}}'_{q,1},{\bf{I}}'_{q,2},...{\bf{I}}'_{q,W}$, 
with expressions as follows.
\begin{equation}\label{eq36}
	\begin{aligned}
		\begin{array}{l}
			{\bf{I}}'_{q,w}={\left| { U''_w } \right|}^2 \left( {\begin{array}{*{20}{c}}
					{\cos(2\pi f_c\frac{2\Delta v_{q,w} T}{C}) }&{\cos(2\pi f_c\frac{4\Delta v_{q,w} T}{C}) }
			\end{array}} \right.\\
			\mathop {\quad \quad \quad \quad \quad \quad\;\left. {\begin{array}{*{20}{c}}
						\cdots &{\cos(2\pi f_c\frac{2\Delta v_{q,w} (N_s-1) T}{C})  }
				\end{array}} \right)}.
		\end{array}
	\end{aligned}
\end{equation}

Finally, the inner product of the vectors ${\bf{I}}'_{q,1},{\bf{I}}'_{q,2},\cdots,{\bf{I}}'_{q,W}$
is calculated, 
which is adopted as the weight of the lattice point $q$, denoted by ${\bf{J}}\left( q \right)$, 
with expression as follows.
\begin{equation}\label{eq37}
	\begin{aligned}
		{\bf{J}}\left( q \right) & = \sum\limits_{k = 1}^{N_s - 1}{\prod\limits_{w = 1}^{W} \left[{\bf{I}}'_{q,w}(k)\right] }  \\
		&	= \sum\limits_{k = 1}^{N_s - 1} {\prod\limits_{w = 1}^{W} \left[{{\left| { U''_w } \right|}^2 \cos \left(2\pi f_c\frac{2\Delta v_{q,w} k T}{C} \right)}\right] }.
	\end{aligned}
\end{equation}

Repeating (\ref{eq35}) $\sim$ (\ref{eq37}), the weights of all lattice points are calculated.
The velocity corresponding to the lattice point with the maximum weight value is adopted as the estimated velocity of target.

\section{Performance Analysis of Multi-BS Cooperative Sensing Algorithm}

This section analyzes the requirements of synchronization accuracy  
and the performance of the proposed symbol-level multi-BS cooperative sensing algorithm.

\subsection{Requirements of Time Synchronization Accuracy}

The synchronization methods used by BSs mainly apply the Global Positioning System (GPS) and IEEE 1588-v2 protocol. 
GPS synchronization realizes the synchronization accuracy of 20$\sim$100 ns,
while the synchronization with IEEE 1588-v2 protocol obtains time synchronization via 
the clock server, with a synchronization accuracy of up to 100 ns \cite{20,21}. 

As for the time synchronization when fusing sensing information from multiple BSs, 
since the vectors of each BS are extracted from a single frame of signal, 
it is necessary to achieve OFDM symbol-level synchronization when fusing the vectors uploaded 
from multiple BSs, 
which requires a time synchronization accuracy 
of $\pm$65 ns for the OFDM symbol-level time alignment among 5G cooperative BSs 
according to the 
report of Zhongxing Telecommunication Equipment (ZTE) Corporation \cite{ZTE}, which could be satisfied by the synchronization level of the 
existing mobile communication systems.

However, as for the similar research area of radar, namely distributed coherent aperture radar, 
the requirement of time synchronization accuracy is much higher. 
Yin \textit{et al.} discovered that 
the time synchronization error should be smaller than 0.028/$B$ 
when the phase synchronization error is 10 degree \cite{18-0}, 
where $B$ is the signal bandwidth. 
When the signal bandwidth is 100 MHz, 
the corresponding time synchronization error 
is smaller than 0.28 ns. 
The symbol-level multi-BS cooperative sensing  
has a much lower requirement of time synchronization 
compared with distributed coherent aperture radar.

In summary, the existing time synchronization methods between multiple BSs 
can fully meet the requirements of symbol-level multi-BS cooperative sensing. 
However, it is difficult to meet the requirements of time synchronization of 
distributed coherent aperture radar, 
so that the symbol-level multi-BS cooperative sensing is more 
suitable to existing mobile communication systems.

\subsection{Comparison of Multi-BS Sensing and Single-BS Sensing}

Since both single-BS sensing and 
multi-BS sensing are related with the matrix ${\bf{B}}_w$ in (\ref{eq6}), 
this subsection analyzes the matrix ${\bf{B}}_w$ for performance analysis.

Assume that the modulus of the signal component in each element 
of matrix ${\bf{B}}_w\,(w=1,2,\cdots,W)$ is $k_w$, 
while the variance of the noise component is $k^2_w\sigma _w^2$. 
The number of OFDM symbols is $N_s$ and the number of subcarriers is $N_c$,
so that the SNR of the result calculated by 2D FFT in (\ref{eq11}) is

\begin{equation}\label{eq38-0}
	\begin{aligned}
		{\bf{SNR}}_{\rm{2D-FFT}} & = \frac{(N_cN_sk_w)^2}{N_cN_sk^2_w\sigma _w^2} \\
		&	=  N_cN_s/\sigma _w^2,
	\end{aligned}
\end{equation}
which is applied to measure the performance of single-BS sensing.

\subsubsection{SNR performance of single-BS sensing}

The SNR performance with the sensing information of single-BS is analyzed. When compressing the matrix ${\bf{B}}_w \,(w=1,2,\cdots,W)$ 
to obtain distance feature vector ${\bf{E}}_w \,(w=1,2,\cdots,W)$ in (\ref{eq13}), 
each element in vector ${\bf{E}}_w$ is the sum of $N_s$ elements in ${\bf{B}}_w$, so that the modulus of the signal component in each element of ${\bf{E}}_w$ is approximately equal to $N_s k_w$, and the variance of the noise component is $N_s k^2_w\sigma _w^2$. 
Similarly, according to (\ref{eq15}), the modulus of the signal component in each element of ${\bf{F}}_w$ is approximately equal to $N_c k_w$, 
while the variance of the noise component is $N_c k^2_w\sigma _w^2$. 

As for the reconstructed vector ${\bf{G}}_w\,(w=1,2,...,W)$ corresponding to the vector ${\bf{E}}_w\,(w=1,2,...,W)$, 
the $k$th element in ${\bf{G}}_w$ is the sum of conjugate multiplication of $(N_c-k)$ pairs of
the elements of ${\bf{E}}_w$ in (\ref{eq22}). 
Considering the noise component, the expression for the $k$-th element of ${\bf{G}}_w$ is 
\begin{equation}\label{eq38}
	\begin{aligned}
		&{\bf{G}}_w(k) \\ =&\frac{ \sum\limits_{a=1}^{N_c-k}[{\bf{E}}_{w,s}(a)+{\bf{E}}_{w,n}(a)] [{\bf{E}}_{w,s}^*(a+k)+{\bf{E}}_{w,n}^*(a+k)]}{N_c-k} \\
		 =& ( N_s k_w )^2  e^{j2\pi k\Delta f\frac{2R_w}{C}}+\frac{\sum\limits_{a=1}^{N_c-k}{\bf{E}}_{w,s}(a){\bf{E}}_{w,n}^*(a+k)}{N_c-k}\\
		&+\frac{\sum\limits_{a=1}^{N_c-k}{\bf{E}}_{w,s}^*(a+k){\bf{E}}_{w,n}(a)}{N_c-k}+\frac{\sum\limits_{a=1}^{N_c-k}{\bf{E}}_{w,n}(a){\bf{E}}_{w,n}^*(a+k)}{N_c-k},
	\end{aligned}
\end{equation}
where ${\bf{E}}_{w,s}$ is the signal component of vector ${\bf{E}}_w$, and ${\bf{E}}_{w,n}$ is the noise component of vector ${\bf{E}}_w$. 
In (\ref{eq38}), only the first term does not contain noise components, while the remaining terms all contain noise components, so that the modulus of the signal component of ${\bf{G}}_w(k)$ is $N^2_s k^2_w$, while the variance of the noise component is $N^2_sk^4_w\sigma _w^2(2N_s+\sigma^2_w)/(N_c-k)$. 

With the DFT in (\ref{eq24}) and removing the imaginary parts of the elements of ${\bf{G}}_{z,w}$ in (\ref{eq25}), the signal component of ${\bf{G}}'_{z,w}(k)\,(k=1,2,...,N_c-1)$ is $N^2_s k^2_w{\cos(2\pi k \Delta f\frac{{2{\Delta R}_{z,w} }}{C}) }$, and the variance of the noise component of ${\bf{G}}'_{z,w}(k)$ is $N^2_sk^4_w\sigma _w^2(N_s+\sigma^2_w/2)/(N_c-k)$. 

If the lattice point $z$ is in the real location of the target, the variance of the sum of the signal component in ${\bf{G}}'_{z,w}$ is
\begin{equation}\label{eq39}
	\begin{array}{l}
		\sigma _s^2 {\rm{ = }}{\left[ {( N_c  - 1) N^2_s k^2_w } \right]} ^2 \\
		\;\;\;\; = (N_c-1)^2 N_s^4 k_w^4,
	\end{array}
\end{equation}
while the variance of the noise component is 
\begin{equation}\label{eq40}
	\begin{array}{l}
		\sigma _{noi}^2 =N^2_sk^4_w\sigma _w^2(N_s+\sigma^2_w/2) {\sum\limits_{k = 1}^{N_c-1} \frac{1}{N_c-k} }
	\end{array}.
\end{equation}

For any positive integer $x$, the following inequality holds.
\begin{equation}\label{eq41}
	\frac{1}{{x + 1}} < \log \left( {1 + \frac{1}{x}} \right).
\end{equation}

Hence, we have
\begin{equation}\label{eq42}
	{\sum\limits_{k = 1}^{N_c-1} \frac{1}{N_c-k} }\\
	<1+\log(N_c-1).
\end{equation}

Thus, the SNR of the sum of elements in ${\bf{G}}'_{z,w}$ is
\begin{equation}\label{eq43}
	\begin{array}{l}
		{{\bf{SNR}}_{z,w}} = \frac{(N_c-1)^2 N_s^4 k_w^4}{N^2_sk^4_w\sigma _w^2(N_s+\sigma^2_w/2) {\sum\limits_{k = 1}^{N_c-1} \frac{1}{N_c-k} }}\\
	    \quad \quad \quad \quad =\frac{(N_c-1)^2 N_s^2 }{\sigma _w^2(N_s+\sigma^2_w/2) {\sum\limits_{k = 1}^{N_c-1} \frac{1}{N_c-k} }}.
	\end{array}
\end{equation}

According to (\ref{eq42}), we have
\begin{equation}\label{eq44}
	\begin{array}{l}
		{{\bf{SNR}}_{z,w}}>\frac{(N_c-1)^2 N_s^2 }{\sigma _w^2(N_s+\sigma^2_w/2) [1+\log(N_c-1)]}.
	\end{array}
\end{equation}

If the SNR of the signal received by BS $w$ is higher than -20 dB, 
$\sigma _w^2$ is not larger than 100. 
Then, we have
\begin{equation}\label{eq45}
	\begin{aligned}
		\begin{array}{ll}
			\frac{(N_c-1)^2 N_s^2 }{\sigma _w^2(N_s+\sigma^2_w/2) [1+\log(N_c-1)]} &\ge \frac{(N_c-1)^2 N_s}{N_c(N_s+50) [1+\log(N_c-1)]} \frac{N_cN_s}{\sigma^2_w}\\
			&> \frac{(N_c-2) N_s}{(N_s+50) [1+\log(N_c-1)]} \frac{N_cN_s}{\sigma^2_w}.\\
		\end{array}	    
	\end{aligned}
\end{equation}

If appropriate parameters are selected (for example, $N_c=128, N_s=256$), we have 
\begin{equation}\label{eq46}
	\begin{aligned}
		\frac{(N_c-2) N_s}{(N_s+50) [1+\log(N_c-1)]}=18.04>>1,   
	\end{aligned}
\end{equation}
so that
\begin{equation}\label{eq47}
	\begin{aligned}
		\begin{array}{ll}
			{\bf{SNR}}_{z,w}>\frac{(N_c-1)^2 N_s^2 }{\sigma _w^2(N_s+\sigma^2_w/2) [1+\log(N_c-1)]}>> \frac{N_cN_s}{\sigma^2_w}
	    \end{array}	    
	\end{aligned}.
\end{equation}

In this case, the SNR of the DFT result of vector ${\bf{G}}_w$ is higher than $N_cN_s/\sigma _w^2 $, which is the SNR of the result calculated by 2D FFT on matrix ${\bf{B}}_w$.

\subsubsection{SNR performance of multi-BS sensing fushion}
The SNR performance with the sensing information fusion of multiple BSs is analyzed. 
For the lattice point $z$, 
all elements of vector ${\bf{G}}'_{z,w}\,(w=1,2,...,W)$ in (\ref{eq25}) are real number, 
and the noise components in the elements follow Gaussian distribution. 
According to the characteristic of Gaussian distribution, 
if $x_1\sim\mathcal{N}(u_1,\delta^2_1)$ and $x_2\sim\mathcal{N}(u_2,\delta^2_2)$ 
are two random variables following Gaussian distribution, 
the product of $x_1$ and $x_2$ follows a Gaussian distribution, 
with the mean and variance being $\frac{u_2 \delta^2_1+u_1 \delta^2_2}{\delta^2_1+\delta^2_2}$ and $\frac{\delta^2_1 \delta^2_2}{\delta^2_1 + \delta^2_2}$, respectively. 
It is proved that 
\begin{equation}\label{eq48}
	\begin{aligned}
		\begin{array}{ll}
			\frac{u_2 \delta^2_1+u_1 \delta^2_2}{\delta^2_1+\delta^2_2}=u_1+\frac{u_2-u_1}{1+\delta^2_2/\delta^2_1}=u_2+\frac{u_1-u_2}{1+\delta^2_1/\delta^2_2}.
		\end{array}	    
	\end{aligned}
\end{equation}

Hence, we have 
\begin{equation}\label{eq49}
	\begin{aligned}
		\begin{array}{ll}
			\min[u_1,u_2]\le\frac{u_2 \delta^2_1+u_1 \delta^2_2}{\delta^2_1+\delta^2_2}\le\max[u_1,u_2]
		\end{array}	    
	\end{aligned}.
\end{equation}

And it is proved that 
\begin{equation}\label{eq50}
	\begin{aligned}
		\left\{\begin{array}{ll}
			\frac{\delta^2_1 \delta^2_2}{\delta^2_1 + \delta^2_2} = \frac{\delta^2_1}{\delta^2_1 + \delta^2_2} \delta^2_2 < \delta^2_2\\
			\frac{\delta^2_1 \delta^2_2}{\delta^2_1 + \delta^2_2} = \frac{\delta^2_2}{\delta^2_1 + \delta^2_2} \delta^2_1 < \delta^2_1
		\end{array}	\right.    
	\end{aligned}.
\end{equation}

Assuming that multiple variables $x_w\,(w=1,2,...,W)$ all follow Gaussian distribution $x_w\sim\mathcal{N}(u_w,\delta^2_w)$, where $\delta^2_1=\delta^2_2=...\delta^2_w$ and $u_1 > u_2>...>u_w$. The mean and variance of $\prod\limits_{w = 1}^W {\mathop x\nolimits_w }$ is $\left( {\sum\limits_{w = 1}^W {\mathop u\nolimits_w } } \right)/W$ and $\delta^2_1/W$. The ratio of the square of the mean of $\prod\limits_{w = 1}^W {x_w}$ to its variance is
\begin{equation}\label{eq50-0}
	\begin{array}{l}
		{{\bf{SNR}}\left(\prod\limits_{w = 1}^W {x_w}\right)} = \frac{\left( {\sum\limits_{w = 1}^W {\mathop u\nolimits_w } } \right)^2/W^2}{\delta^2_1/W}\\
		\quad \quad \quad \quad \quad \quad \quad \;=\frac{\left( {\sum\limits_{w = 1}^W {\mathop u\nolimits_w } } \right)^2/W }{\delta^2_1}.
	\end{array}
\end{equation}
Therefore, if $\sum\limits_{w = 1}^W {\mathop u\nolimits_w }>\sqrt W u_1$, then the ratio of the square of the mean of $\prod\limits_{w = 1}^W {x_w}$ to its variance is higher than that of $x_1$, whose ratio of mean square to variance is the largest.

According to (\ref{eq26}), the inner product of the vectors ${\bf{G}}'_{z,1},{\bf{G}}'_{z,2},\cdots,{\bf{G}}'_{z,W}$ can be regarded as 
the sum of the elements of a vector $\textbf{L}_z$, whose $k$-th element is
\begin{equation}\label{eq51}
	\begin{aligned}
		{\bf{L}}_z(k) & = {\prod\limits_{w = 1}^{W} \left[{\bf{G}}'_{z,w}(k)\right] }  \\
		&= {\prod\limits_{w = 1}^{W} \left[{{\left| { U'_w } \right|}^2 \cos \left( {2\pi k\Delta f\frac{{2 {\Delta R}_{z,w} }}{C}} \right)}\right] }.
	\end{aligned}
\end{equation}

Since all elements in the vectors 
${\bf{G}}'_{z,1}, {\bf{G}}'_{z,2},\cdots,{\bf{G}}'_{z,W}$ contain noise components that follow Gaussian distribution, ${\bf{L}}_z(k)$ also follows Gaussian distribution. If multiple received signals have the similar SNR, then the SNR of the result of multi-BS sensing fushion
is higher than the SNR of the result from single-BS sensing.
Thus, in terms of location estimation,
multi-BS cooperative sensing has better sensing performance than single-BS sensing. 

In terms of velocity estimation,
the similar derivations on the SNR performance can be performed to prove the performance improvement of multi-BS cooperative sensing.

\section{Simulation Results}

The proposed symbol-level multi-BS cooperative sensing method is evaluated in this section. 
Table \ref{tab2} \cite{19} provides the main simulation parameters.
The cooperative sensing algorithm undergoes 1000 times Monte Carlo simulations. 

\begin{table}[!t]
	\renewcommand\arraystretch{1.5}
	\caption{\label{tab2}Key simulation parameters}
	\begin{center}
		\begin{tabular}{c c c}
			\hline
			\hline
			Symbol & Meaning & Value/Range \\
			\hline
			$f_c$  & Frequency of band signal & 24 GHz    \\ \hline
			$B$ & Signal bandwidth & 93.1 MHz   \\ \hline
			$T$ & Full duration of OFDM symbol & 12.375 $\mu \rm{s}$   \\ \hline
			$N_c$ & Number of subcarriers & 128    \\ \hline
			$N_s$   & Number of OFDM 	symbols & 256    \\ \hline
			SNR  & SNR of the received echo signal & --5 $\sim$ --20~$\rm{dB}$    \\ \hline  
			$(x_t,y_t)$  & Location of target & (0,0) $\sim$ (10,10)    \\ \hline
			$V$  & Velocity of target & 27 m/s    \\ \hline
			$W$  & Number of BSs & 2 $\sim$ 4    \\ 
			\hline
			\hline
		\end{tabular}
	\end{center}
\end{table}

\subsection{Single-BS Signal Preprocessing}

\begin{figure}[!ht]
	\centering
	\includegraphics[width=0.48\textwidth]{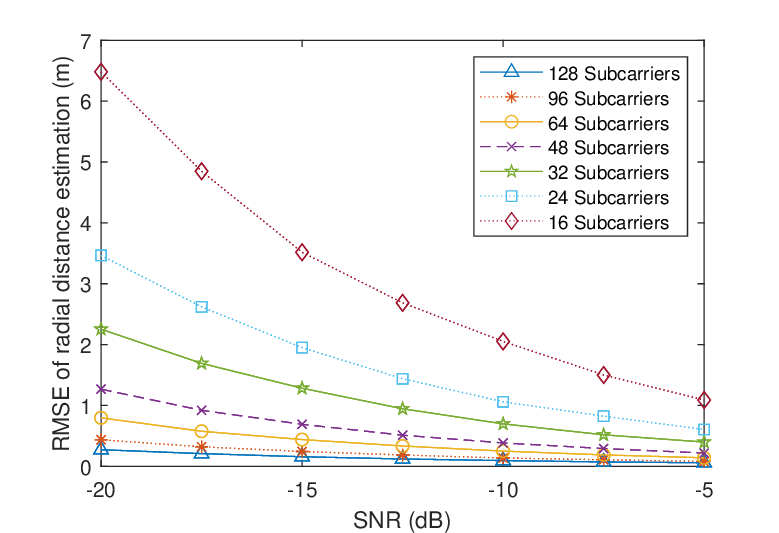}
	\caption{RMSE of radial distance versus SNR.}
	\label{fig6}
\end{figure}

\begin{figure}[!ht]
	\centering
	\includegraphics[width=0.48\textwidth]{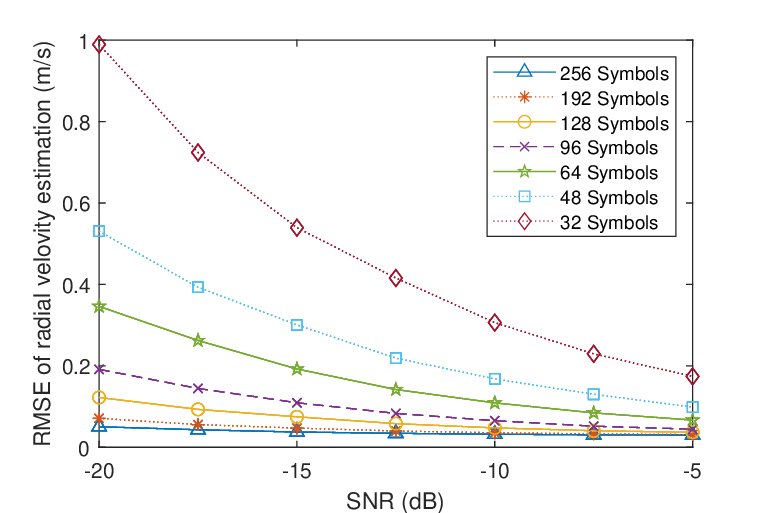}
	\caption{RMSE of radial velocity versus SNR.}
	\label{fig7}
\end{figure}

As shown in Figs. \ref{fig6} and \ref{fig7}, 
the root-mean-squared errors (RMSEs) are decreasing when the amount of time-frequency resources
or SNR is increasing.

However, in the regimes of high SNR, 
the RMSEs with different number of subcarriers or 
symbols decrease slightly when the SNR is increasing, 
because the main factor causing the sensing errors 
is the interval of sample points when estimating velocity and distance in the regimes of high SNR.

\subsection{Multi-BS Sensing Information Fusion}

\subsubsection{Impact of the locations of BSs on sensing accuracy}

\begin{figure}[!ht]
	\centering
	\includegraphics[width=0.48\textwidth]{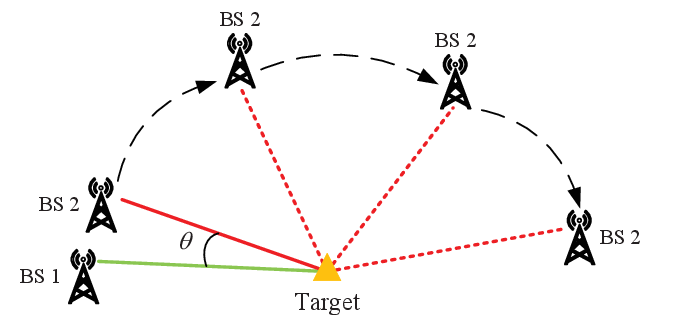}
	\caption{The angle between two BSs.}
    \label{fig6a}
\end{figure}

\begin{figure}[!ht]
	\centering
	\includegraphics[width=0.48\textwidth]{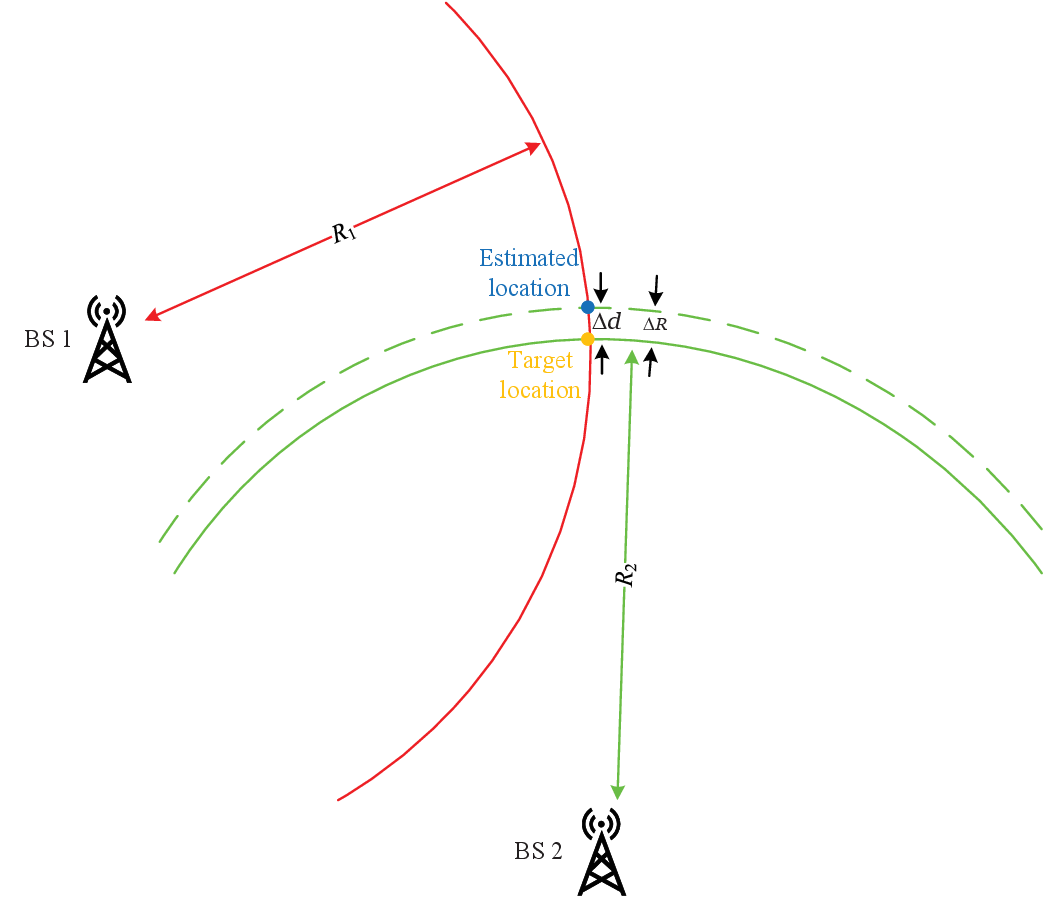}
	\caption{The angle between two BSs and target is close to 90 degree.}
	\label{fig6b}
\end{figure}

\begin{figure}[!ht]
	\centering
	\includegraphics[width=0.48\textwidth]{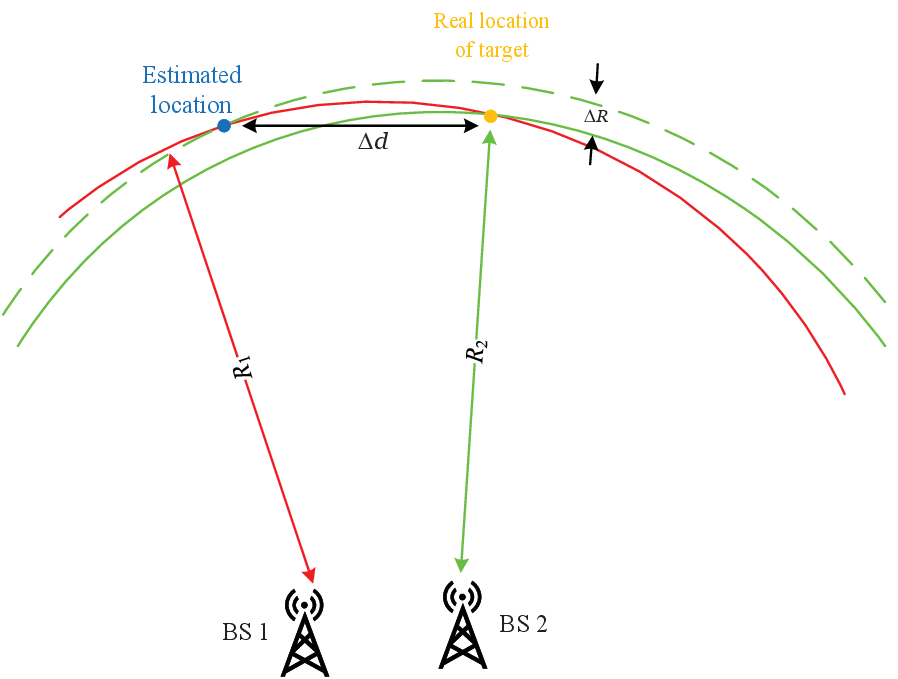}
	\caption{The angle between two BSs and target is far from 90 degree.}
	\label{fig6c}
\end{figure}

\begin{figure}[!ht]
	\centering
	\includegraphics[width=0.48\textwidth]{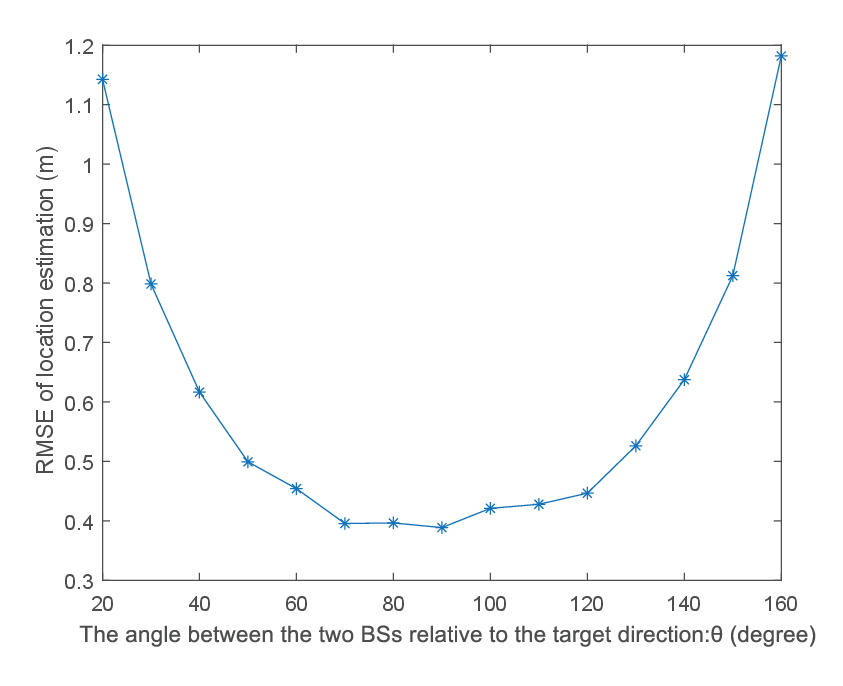}
	\caption{RMSE of location estimation versus the angle between two BSs.}
	\label{fig9}
\end{figure}

The impact of the locations of multiple BSs on the sensing accuracy is revealed. 
We simulate the RMSE of location estimation 
with the angle between two BSs $\theta$ changing from 20 degrees to 160 degrees, as shown in Fig. \ref{fig6a}. 
As shown in Fig. \ref{fig9}, the RMSE of location estimation with symbol-level cooperative sensing 
is the minimum when the angle between two BSs is 90 degrees. 
The reason is as follows. 
As shown in Fig. \ref{fig6b}, when $\theta$ is close to 90 degree, 
the error of the distance estimation by BS $1$ has a small impact on the 
estimated location of target. 
However, when $\theta$ is far from 90 degree, 
as shown in Fig. \ref{fig6c}, 
the error of the distance estimation by BS $1$ has a large impact on the 
estimated location of target.

\subsubsection{Impact of the number of BSs on sensing accuracy}

\begin{figure}[!ht]
	\centering
	\includegraphics[width=0.48\textwidth]{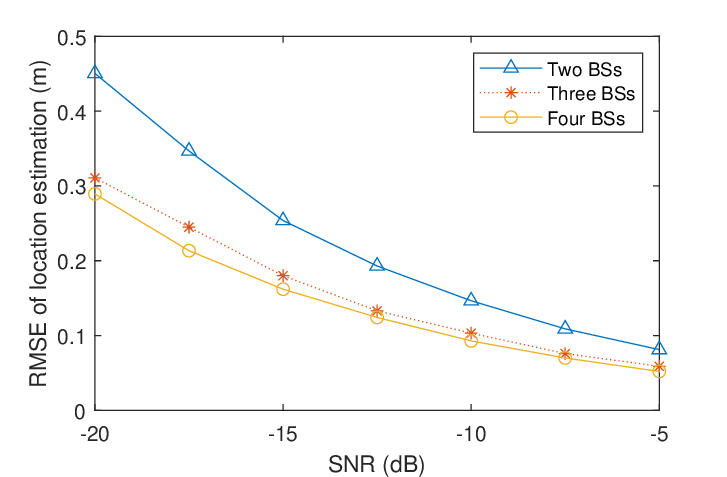}
	\caption{RMSE of location estimation versus SNR.}
	\label{fig10}
\end{figure}

\begin{figure}[!ht]
	\centering
	\includegraphics[width=0.48\textwidth]{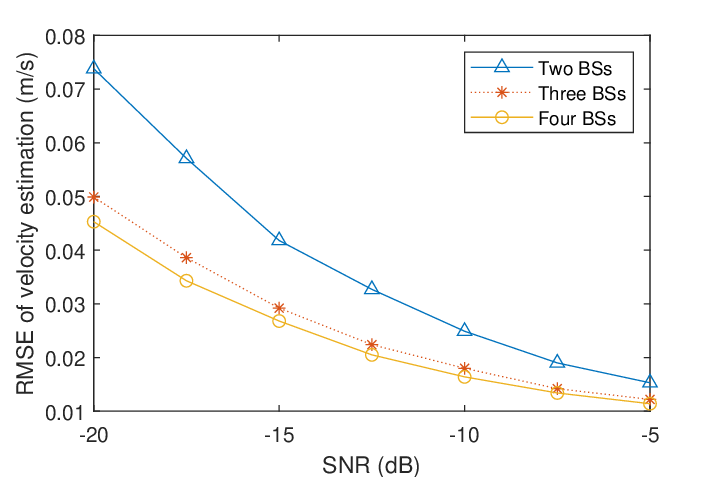}
	\caption{RMSE of velocity estimation versus SNR.}
	\label{fig11}
\end{figure}

Figs. \ref{fig10} and \ref{fig11} show the accuracy of 
location and velocity estimation with different number of BSs and SNR. 
It is shown that the accuracy of location and velocity estimation is significantly improving 
when we increase the number of BSs or SNR.

\subsubsection{Comparison between single-BS sensing and multi-BS sensing }
\begin{figure}[!ht]
	\centering
	\includegraphics[width=0.48\textwidth]{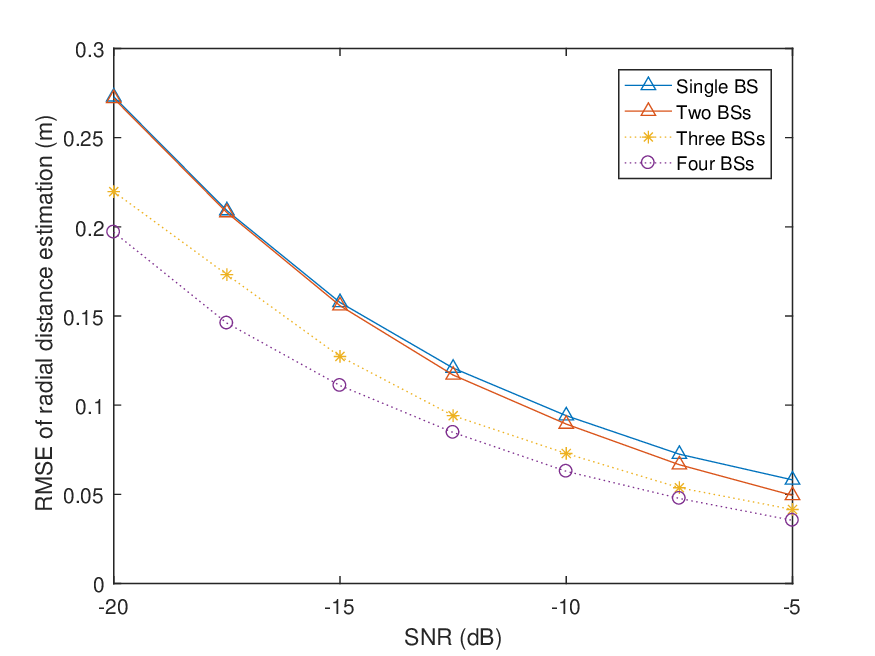}
	\caption{RMSE of radial distance versus SNR with different number of BSs.}
	\label{fig12}
\end{figure}

\begin{figure}[!ht]
	\centering
	\includegraphics[width=0.48\textwidth]{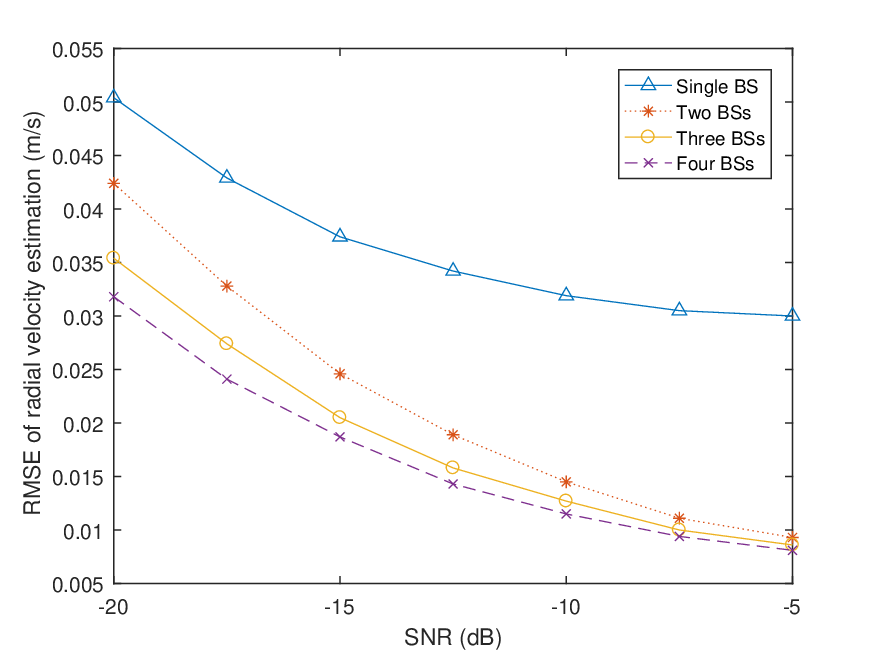}
	\caption{RMSE of radial velocity versus SNR with different number of BSs.}
	\label{fig13}
\end{figure}

The RMSEs of distance and radial velocity estimation with multi-BS cooperative sensing are 
much lower than those with single-BS sensing, as illustrated in Figs. \ref{fig12} and \ref{fig13}.
Specifically, we discover that the performance improvement for radial velocity estimation 
is more significant compared with the performance improvement for 
distance estimation.
The reason is as follows. 
The radial velocity is the projection of the velocity of target.
The proposed symbol-level multi-BS cooperative sensing method
improves the accuracy of radial velocity estimation, 
which has a greater improvement 
on the velocity estimation accuracy.

\subsubsection{Comparison between symbol-level and data-level cooperative sensing}
The symbol-level multi-BS cooperative sensing method is verified. 
We compared the symbol-level cooperative sensing algorithm with the 
data-level cooperative sensing algorithm, 
namely a probabilistic Maximum Likelihood Estimation (MLE) algorithm in \cite{12}. 
The MLE algorithm calculates a posterior likelihood function based on 
the estimation results of each BS, 
the distance and radial velocity of target calculated by the 2D FFT algorithm of 
each BS are used as the input parameters of MLE algorithm 
in the simulation of the MLE algorithm.

When the sensing information fusion algorithm is 
used to calculate the location and velocity of target, 
the likelihood functions of all lattice points are calculated iteratively, 
and the location of the lattice point with 
maximum likelihood function is taken as the estimation location of target. 
Taking lattice point $z$ for an example, 
if the estimated distance between lattice point $z$ and BS $w$ is $R'_w$, 
the distance between this lattice and BS $w$ is $R_{z,w}$, 
the power of noise in the signal received by BS $w$ is $\sigma _w^2$, 
then the conditional probabilities are
\begin{equation}\label{eq52}
	\begin{aligned}
		P(R'_w/R_z,w)=\frac{1}{{\sqrt {2\pi \mathop \sigma \nolimits_w^2 } }}\mathop e\nolimits^{- \frac{{ \mathop {\left( {\mathop R\nolimits_z^'  - \mathop R\nolimits_{z,w} } \right)}\nolimits^2 }}{{2\mathop \sigma \nolimits_w^2 }}}, w = 1,2, \cdots ,W.
	\end{aligned}
\end{equation}

Due to the dispersed distribution of multiple BSs, 
the random variables $R'_w, w = 1, 2,\cdots, W$ are independent, 
and the posterior likelihood function $L(z)$ is the joint probability distribution function of all the random variables $R'_w, w = 1, 2,\cdots, W$.
\begin{equation}\label{eq53}
	\begin{aligned}
		L(z)=\prod\limits_{w = 1}^W {{\bf{P}}(R'_w/R_z,w)}
	\end{aligned}.
\end{equation}

Since the locations of lattice points are close to the real location of target 
and the number of lattice points is not large, 
the posterior likelihood functions of all lattice points are traversed
to find the lattice point with the maximum likelihood function,
whose location is the estimation of the location of target. 
Similarly, the velocity estimation method using the MLE algorithm
can be designed.

\begin{figure}[!ht]
	\centering
	\includegraphics[width=0.48\textwidth]{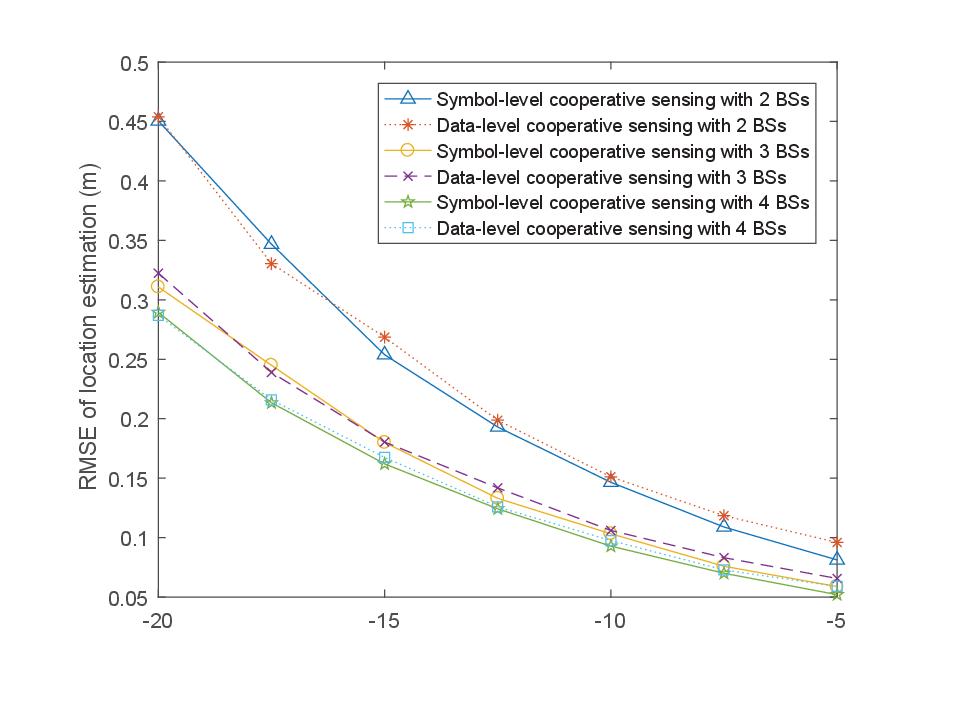}
	\caption{RMSE of location estimation versus SNR with symbol-level and data-level multi-BS cooperative sensing.}
	\label{fig14}
\end{figure}

\begin{figure}[!ht]
	\centering
	\includegraphics[width=0.48\textwidth]{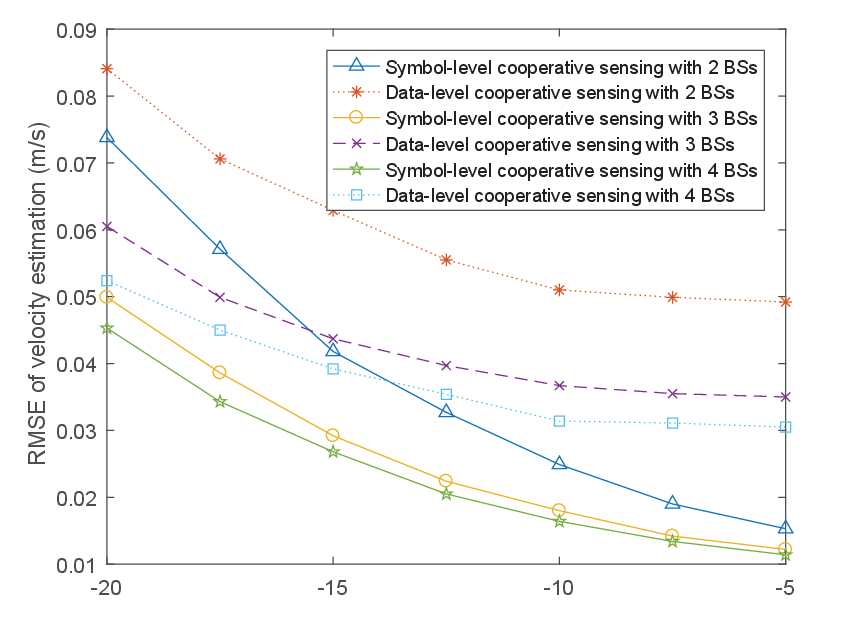}
	\caption{RMSE of velocity estimation versus SNR with symbol-level and data-level multi-BS cooperative sensing.}
	\label{fig15}
\end{figure}

As depicted in Fig. \ref{fig14}, 
the RMSEs of location estimation with symbol-level cooperative sensing algorithm are lower than those with 
data-level cooperative sensing algorithm. 
Besides, the performance improvement is more obvious
as the SNR increases.
In addition, the symbol-level cooperative sensing algorithm has a more significant performance improvement 
in velocity estimation compared with that in location estimation, as shown in Fig. \ref{fig15}.

\section{Conclusions}

A symbol-level multi-BS cooperative sensing method suitable to mobile communication systems is proposed in this paper, reducing the requirement of multi-BS synchronization compared with signal-level cooperative sensing. The cooperative sensing method includes single-BS signal preprocessing and multi-BS information fusion. It is evaluated that symbol-level multi-BS cooperative sensing has better sensing performance compared with single-BS sensing and data-level multi-BS cooperative sensing based on ML estimation. 
Although the multi-BS cooperative sensing method in this paper is designed for mobile communication system with OFDM signal, it can also be applied in multi-radar cooperative sensing system with the vectors containing the distance and velocity information of target extracted from the received echo signal.
This paper provides a guideline for the design of multi-BS cooperative sensing method exploiting the networked mobile communication system.
Notice that ISAC system combined with non-orthogonal multiple access (NOMA) 
has advances in the maximum allowable number of users and the 
spectrum utilization \cite{future}. Hence, the 
ISAC enabled multi-BS cooperative sensing with NOMA is the future work.

\bibliographystyle{IEEEtran} 
\bibliography{reference}

\begin{IEEEbiography}[{\includegraphics[width=1.1in,height=1.4in,clip,keepaspectratio]{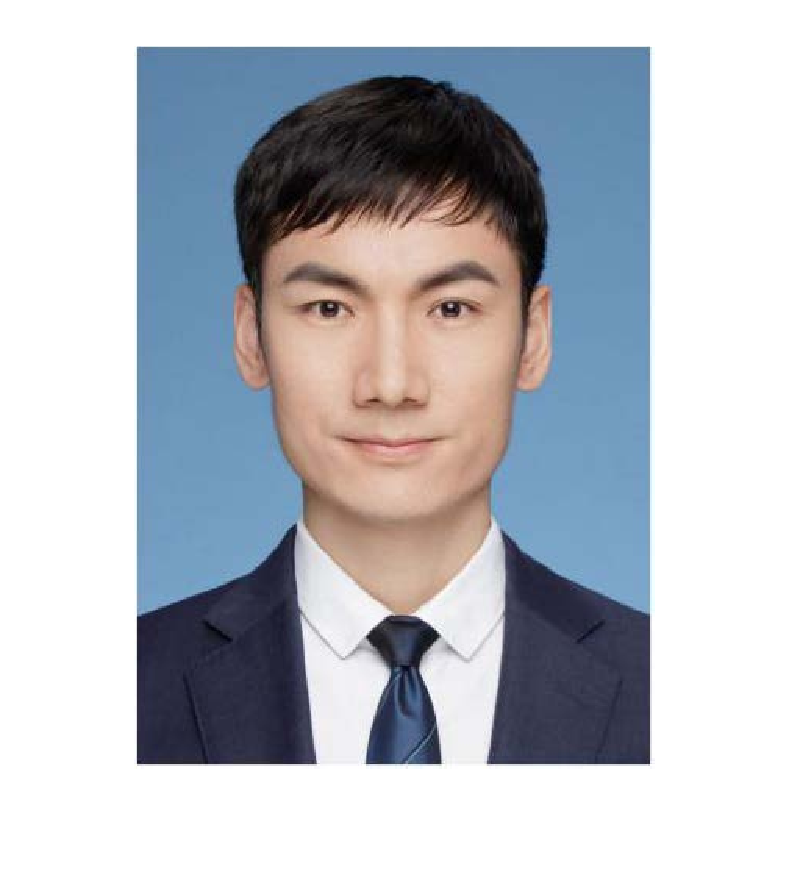}}]
	{Zhiqing Wei}
	(S'12-M'15) received his B.E. and Ph.D. degrees from BUPT in 2010 and 2015. Now he is an associate professor at BUPT. He was granted the Exemplary Reviewer of IEEE Wireless Communications Letters in 2017, the Best Paper Award of International Conference on Wireless Communications and Signal Processing 2018. He was the Registration Co-Chair of IEEE/CIC International Conference on Communications in China (ICCC) 2018 and the publication Co-Chair of IEEE/CIC ICCC 2019. His research interest is the performance analysis and optimization of mobile ad hoc networks.
\end{IEEEbiography}

\begin{IEEEbiography}[{\includegraphics[width=1in,height=1.25in,clip,keepaspectratio]{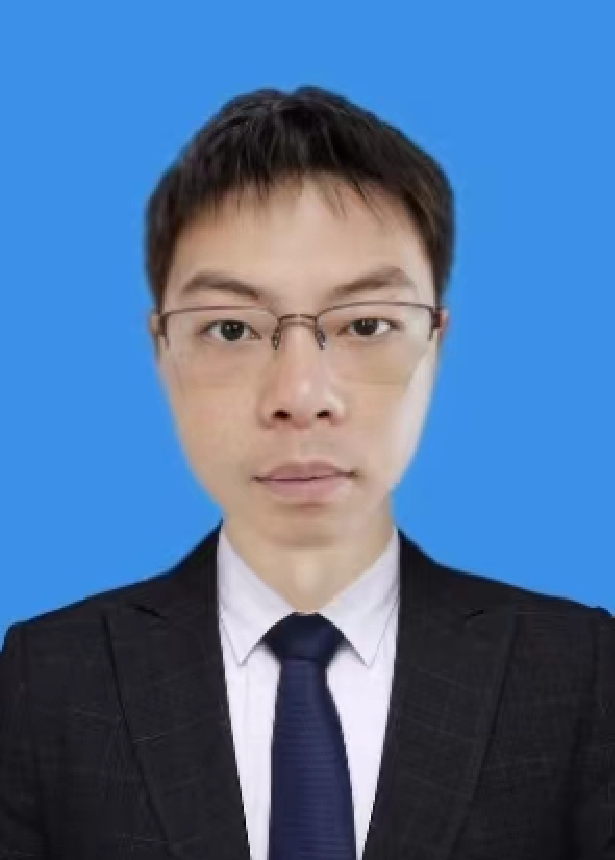}}]{Ruizhong Xu}
	received the M.S. degree in School of Information and Communication Engineering, Beijing University of Posts and Telecommunications (BUPT) in 2023. He is currently pursuing his Ph.D. degree with Beijing University of Posts and Telecommunication (BUPT). His research interests include integrated sensing and communication and cooperative sensing.
\end{IEEEbiography}

\begin{IEEEbiography}[{\includegraphics[width=1.1in,height=1.4in,clip,keepaspectratio]{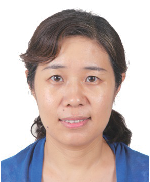}}]{Zhiyong Feng}
	(M'08-SM'15) received her B.S., M.S., and Ph.D. degrees from BUPT, Beijing, China. She is a Professor with the School of Information and Communication Engineering, BUPT, and the director of the Key Laboratory of Universal Wireless Communications, Ministry of Education, China. Her research interests include wireless network architecture design and radio resource management in 5th generation mobile networks (5G), spectrum sensing and dynamic spectrum management in cognitive wireless networks, universal signal detection and identification, and network information theory. She is a senior member of IEEE and active in standards development, such as ITU-R WP5A/5C/5D, IEEE 1900, ETSI, and CCSA.
\end{IEEEbiography}

\begin{IEEEbiography}[{\includegraphics[width=1.25in,height=1.4in,clip,keepaspectratio]{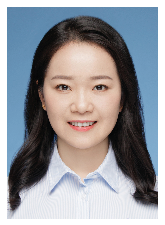}}]{Huici Wu} (Member, IEEE) received the Ph.D degree from Beijing University of Posts and Telecommunications (BUPT), Beijing, China, in 2018. From 2016 to 2017, she visited the Broadband Communications Research (BBCR) Group, University of Waterloo, Waterloo, ON, Canada. She is now an Associate Professor at BUPT. Her research interests are in the area of wireless communications and networks, with current emphasis on collaborative air-to-ground communication and wireless access security.
\end{IEEEbiography}

\begin{IEEEbiography}[{\includegraphics[width=1.1in,height=1.25in,clip,keepaspectratio]{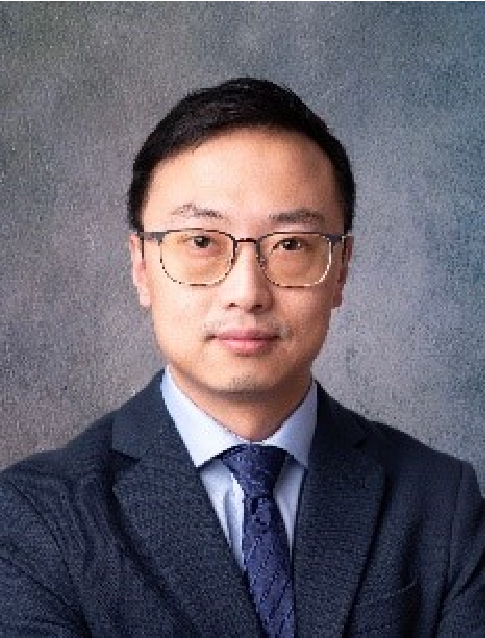}}]{Ning Zhang} (Senior Member, IEEE)  received the Ph.D degree in Electrical and Computer Engineering from University of Waterloo, Canada, in 2015. After that, he was a postdoc research fellow at University of Waterloo and University of Toronto, respectively. Since 2020, he has been an Associate Professor in the Department of Electrical and Computer Engineering at University of Windsor, Canada. His research interests include connected vehicles, mobile edge computing, wireless networking, and security. He is a Highly Cited Researcher (Web of Science). He serves/served as an Associate Editor of IEEE Transactions on Mobile Computing, IEEE Communications Surveys and Tutorials, IEEE Internet of Things Journal, and IEEE Transactions on Cognitive Communications and Networking. He also serves/served as a TPC chair for IEEE VTC 2021 and IEEE SAGC 2020, a general chair for IEEE SAGC 2021, a chair for track of several international conferences and workshops including IEEE ICC, VTC, INFOCOM Workshop, and Mobicom Workshop. He received a number of Best Paper Awards from conferences and journals, such as IEEE Globecom, IEEE ICC, IEEE ICCC, IEEE WCSP, and Journal of Communications and Information Networks. He also received IEEE TCSVC Rising Star Award and IEEE ComSoc Young Professionals Outstanding Nominee Award. He serves as the Vice Chair for IEEE Technical Committee on Cognitive Networks and IEEE Technical Committee on Big Data.
\end{IEEEbiography}

\begin{IEEEbiography}[{\includegraphics[width=1in,height=1.25in,clip,keepaspectratio]{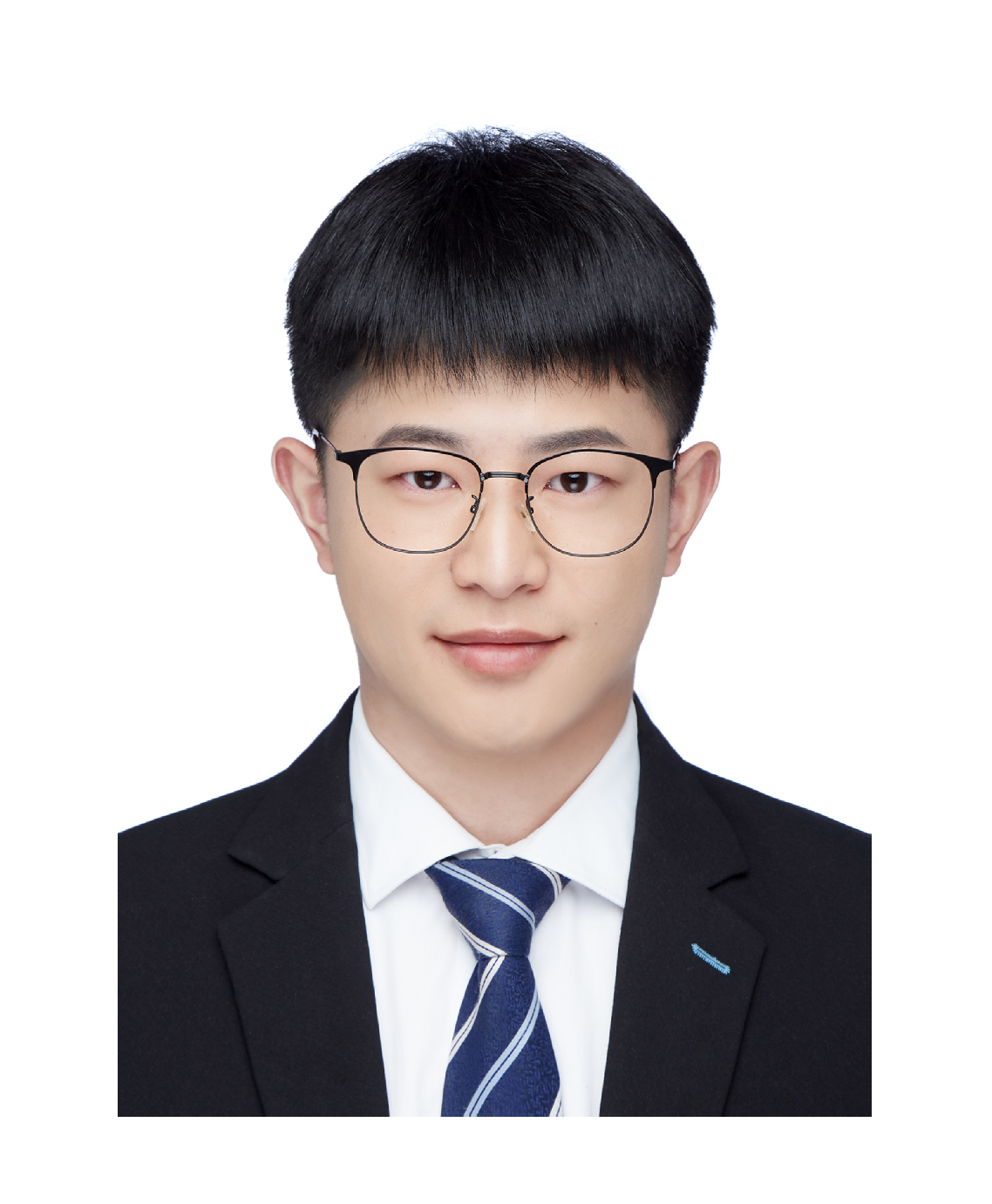}}]{Wangjun Jiang}
	received the B.S. degree in School of Electronic and Information Engineering, Beijing Jiaotong University (BJTU) in 2019. He is currently pursuing his Ph.D. degree with Beijing University of Posts and Telecommunication (BUPT). His research interests include integrated sensing and communication and network sensing.
\end{IEEEbiography}
\begin{IEEEbiography}[{\includegraphics[width=1in,height=1.25in,clip,keepaspectratio]{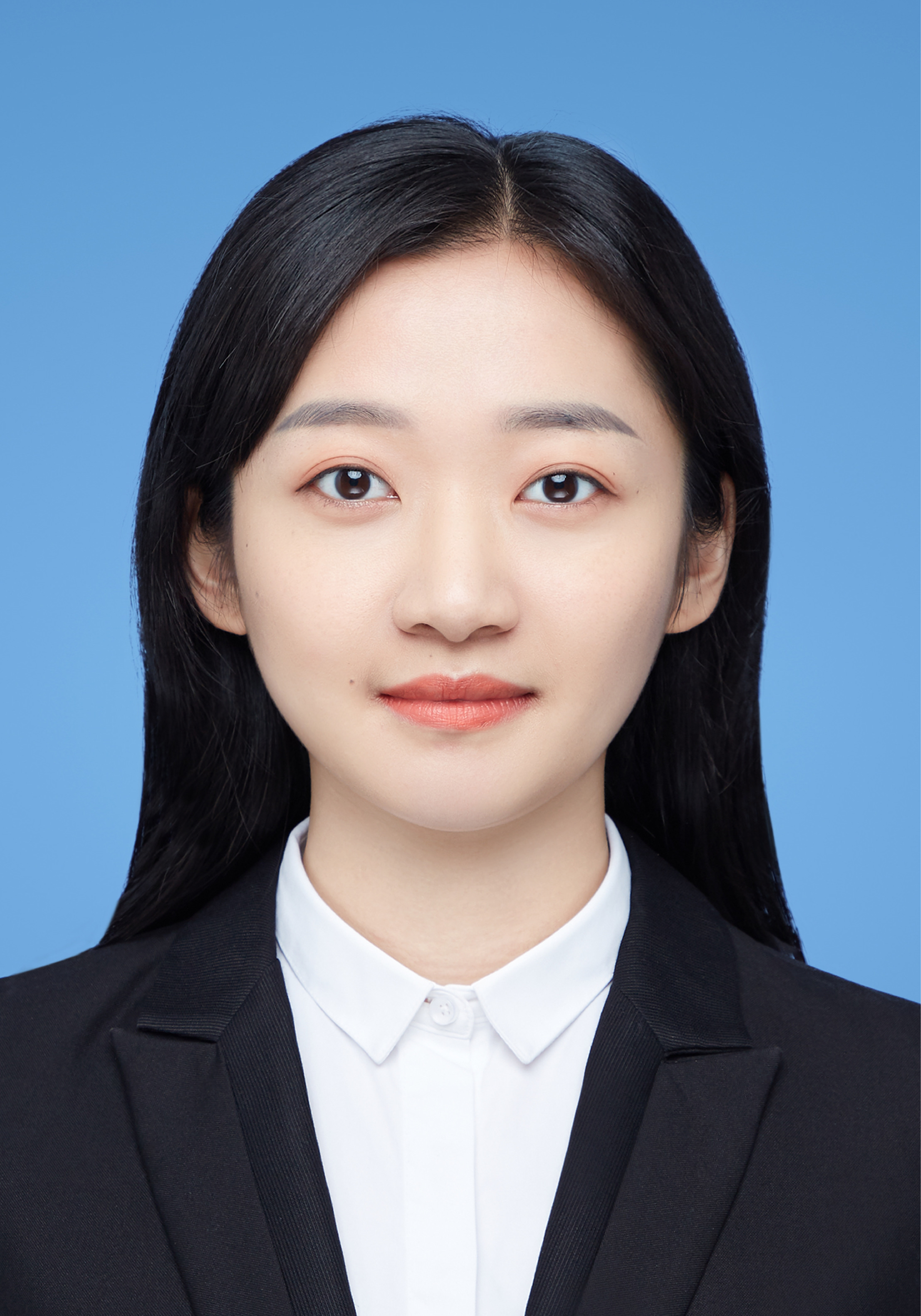}}]{Xiaoyu Yang}
	received the M.S. degree in information and communication engineering from the Beijing University of Posts and Telecommunications, Beijing, China, in 2022. She is currently pursuing the Ph.D. degree with the School of Information and Communication Engineering, Beijing University of Posts and Telecommunications, Beijing, China. Her research interests include integrated sensing and communications and reconfigurable intelligent surface.
\end{IEEEbiography}

\ifCLASSOPTIONcaptionsoff
\newpage
\fi

\end{document}